\documentclass[twocolumn]{aastex61}
\hypersetup{citecolor=cyan,linkcolor=blue}

\usepackage{gensymb}

\shorttitle{[C{\sc ii}] in CR7}
\shortauthors{Matthee et al.}

\begin{document}

\title{ALMA reveals metals yet no dust within multiple components in CR7}

\correspondingauthor{J. Matthee}
\email{matthee@strw.leidenuniv.nl}

\author[0000-0003-2871-127X]{J. Matthee}
\affil{Leiden Observatory, Leiden University, PO\ Box 9513, NL-2300 RA, Leiden, The Netherlands}

\author{D. Sobral}

\affiliation{Department of Physics, Lancaster University, Lancaster, LA1 4YB, UK}
\affil{Leiden Observatory, Leiden University, PO\ Box 9513, NL-2300 RA, Leiden, The Netherlands}

\author{F. Boone}
\affiliation{Universit\'e de Toulouse; UPS-OMP; IRAP; Toulouse, France}

\author{H. R\"ottgering}
\affiliation{Leiden Observatory, Leiden University, PO\ Box 9513, NL-2300 RA, Leiden, The Netherlands}

\author{D. Schaerer}
\affiliation{Observatoire de Gen\`eve, Universit\' e Gen\`eve, 51 Ch. des Maillettes, 1290 Versoix, Switzerland}
\affiliation{Universit\'e de Toulouse; UPS-OMP; IRAP; Toulouse, France}

\author{M. Girard}
\affiliation{Observatoire de Gen\`eve, Universit\' e Gen\`eve, 51 Ch. des Maillettes, 1290 Versoix, Switzerland}

\author{A. Pallottini}
\affiliation{Centro Fermi, Museo Storico della Fisica e Centro Studi e Ricerche ``Enrico Fermi'', Piazza del Viminale 1, Roma, 00184, Italy}
\affiliation{Cavendish Laboratory, University of Cambridge, 19 J. J. Thomson Ave., Cambridge CB3 0HE, UK}
\affiliation{Kavli Institute for Cosmology, University of Cambridge, Madingley Road, Cambridge CB3 0HA, UK}
\affiliation{Scuola Normale Superiore, Piazza dei Cavalieri 7, I-56126 Pisa, Italy}

\author{L. Vallini}
\affiliation{Nordita, KTH Royal Institute of Technology and Stockholm University, Roslagstullsbacken 23, SE-10691 Stockholm, Sweden}

\author{A. Ferrara}
\affiliation{Scuola Normale Superiore, Piazza dei Cavalieri 7, I-56126 Pisa, Italy}
\affiliation{Kavli IPMU, The University of Tokyo, 5-1-5 Kashiwanoha, Kashiwa 277-8583, Japan}

\author{B. Darvish}
\affiliation{Cahill Center for Astrophysics, California Institute of Technology, 1216 East California Boulevard, Pasadena, CA 91125, USA}

\author{B. Mobasher}
\affiliation{Department of Physics and Astronomy, University of California, Riverside, CA 92521, USA}



\begin{abstract}

\noindent We present spectroscopic follow-up observations of CR7 with ALMA, targeted at constraining the infrared (IR) continuum and [C{\sc ii}]$_{158 \mu \rm m}$ line-emission at high spatial resolution matched to the {\it HST}/WFC3 imaging. CR7 is a luminous Ly$\alpha$ emitting galaxy at $z=6.6$ that consists of three separated UV-continuum components. Our observations reveal several well-separated components of [C{\sc ii}] emission. The two most luminous components in [C{\sc ii}] coincide with the brightest UV components (A and B), blue-shifted by $\approx 150$ km s$^{-1}$ with respect to the peak of Ly$\alpha$ emission. Other [C{\sc ii}] components are observed close to UV clumps B and C and are blue-shifted by $\approx300$ and $\approx80$ km s$^{-1}$ with respect to the systemic redshift. We do not detect FIR continuum emission due to dust with a 3$\sigma$ limiting luminosity L$_{\rm IR} (T_d = 35 \rm \, K) < 3.1\times10^{10}$ L$_{\odot}$. This allows us to mitigate uncertainties in the dust-corrected SFR and derive SFRs for the three UV clumps A, B and C of 28, 5 and 7 M$_{\odot}$ yr$^{-1}$. All clumps have [C{\sc ii}] luminosities consistent within the scatter observed in the local relation between SFR and L$_{\rm [CII]}$, implying that strong Ly$\alpha$ emission does not necessarily anti-correlate with [C{\sc ii}] luminosity. Combining our measurements with the literature, we show that galaxies with blue UV slopes have weaker [C{\sc ii}] emission at fixed SFR, potentially due to their lower metallicities and/or higher photoionisation. Comparison with hydrodynamical simulations suggests that CR7's clumps have metallicities of $0.1<\rm Z/Z_{\odot}<0.2$. The observed ISM structure of CR7 indicates that we are likely witnessing the build up of a central galaxy in the early Universe through complex accretion of satellites.  
\end{abstract}


\section{Introduction} \label{sec:intro}
Characterising the properties of the interstellar medium (ISM) of the first generations of galaxies is one of the prime goals of observational astrophysics. With the advent of the Atacama Large Millimetre Array (ALMA), direct measurements of the ISM are now becoming possible for typical star-forming galaxies in the early Universe ($z>6$, e.g. \citealt{Maiolino2015,Watson2015,Knudsen2017}), in addition to studies of bright quasar hosts and galaxies with extreme bursts of star-formation \citep[e.g.][]{Swinbank2012,Riechers2013,Wang2013,Decarli2017,Riechers2017}. These measurements are very valuable in constraining models of early galaxy formation \citep[e.g.][]{Ceverino2010,Hopkins2014,Pallottini2017}.

Rest-frame far-infrared continuum measurements (redshifted to sub-millimetre wavelengths detectable by ALMA) can provide a direct determination of the dust mass, temperature and attenuation \citep[e.g.][]{daCunha2015,Bouwens2016,Scoville2016,Scoville2017}, and constrain the rate at which dust has been produced \citep[e.g.][]{Hirashita2014,Michalowski2015,Mancini2015,Mancini2016}. Combining measurements of emission lines in the far-infrared such as the fine-structure lines [C{\sc ii}]$_{158 \mu \rm m}$, [NII]$_{205 \mu \rm m}$ and [OIII]$_{88 \mu \rm m}$ allows to constrain the star formation rate (SFR), metallicity, density and ionisation state of the gas, both at low redshift \citep[e.g.][]{Ferkinhoff2010,DeLooze2014,Herrera-Camus2015} and at high-redshift  \citep[e.g.][]{Inoue2016,Carniani2017}. Furthermore, spatially resolved emission-line measurements can probe the dynamical structure of the gas \citep[e.g.][]{Jones2017,Smit2017}.

Besides quasar-hosts, dust continuum at $z>6$ has been detected in a few sources \citep{Watson2015,Laporte2017}, but most sources show very little dust \citep[e.g.][]{Schaerer2015,Bouwens2016}. Observations of the [C{\sc ii}] fine-structure cooling line indicate a large scatter in [C{\sc ii}] luminosities at fixed SFR, particularly when compared to estimates based on the UV \citep[e.g.][]{DeLooze2014}. While some sources have similar [C{\sc ii}] luminosities as those on the local relation between SFR(UV) and L$_{\rm [CII]}$, \citep[e.g.][]{Capak2015,Smit2017,Jones2017}, others have significantly fainter [C{\sc ii}] luminosities at fixed UV SFR \citep[e.g.][]{Ouchi2013,Ota2014}. A potential explanation is that this is due to a selection bias. A large number of early sources observed by the ALMA were selected based on their Ly$\alpha$ emission needed to measure redshifts. Strong Ly$\alpha$ emission is typically associated with a high ionisation state and/or lower metallicity \citep[e.g.][]{Nakajima2016,Trainor2016,Matthee2017GALEX,Stark2017}, which should result in a deficit in [C{\sc ii}] luminosity due to photo-dissociation \citep[e.g.][]{Vallini2015}. This can be tested with deeper [C{\sc ii}] observations of sources that span a wider parameter space \citep[e.g.][]{Knudsen2017,Bradac2017}.

Here we present deep spectroscopic observations of the COSMOS Redshift 7 galaxy (CR7) with ALMA, targeting the far-infrared dust continuum emission and the [C{\sc ii}] line. CR7 has been identified as the most luminous Ly$\alpha$ emitter (LAE) at $z=6.6$ based on narrow-band imaging with Subaru/Suprime-Cam \citep{Matthee2015}, with L$_{\rm Ly\alpha} = 8.5\times10^{43}$ erg s$^{-1}$ and EW$_{0, \rm Ly\alpha} = 210$ {\AA} \citep{Sobral2015}. Independently, \cite{Bowler2012,Bowler2014} identified CR7 as a candidate luminous Lyman-break galaxy at $z\sim6-7$ with ground-based near-infrared imaging. Hence, CR7 is also among the most UV-luminous galaxies known at $z\sim7$, with $M_{1500} = -22.2\pm0.1$ (see \citealt{Matthee2017spec} for even more luminous ones).

Besides strong Ly$\alpha$ emission, \cite{Sobral2015} identified a narrow He{\sc ii} emission-line in near-infrared spectroscopic follow-up observations, and obtained limits on UV metal lines such as C{\sc iv} and C{\sc iii}]. {\it HST}/WFC3 near-infrared imaging revealed three separate rest-frame UV components, of which the brightest component (A) is closest to the peak of Ly$\alpha$ surface brightness on which spectroscopic observations were centred. The other two components (B and C) have photometric redshifts consistent with $z\gtrsim6.5$. 

These properties led to spectacular interpretations; clump A could contain low metallicity, hot (T$_{\rm eff} \gtrsim 10^5$K) PopIII-like stars \citep[e.g.][]{Sobral2015,Visbal2016} or a direct collapse black hole \citep[e.g.][]{Pallottini2015,Dijkstra2016,Hartwig2016,Agarwal2016,Smidt2016,Smith2016,Agarwal2017}. However, in independent analyses, \cite{Bowler2016} presented evidence of the possible presence of [O{\sc iii}] line-emission inferred from deblended {\it Spitzer}/IRAC photometry. While the \cite{Bowler2016} measurements are still consistent with a direct collapse or low metallicity AGN \citep[e.g.][]{Agarwal2017,Pacucci2017}, the lower claimed significance of the He{\sc ii} line by \cite{Shibuya2017} removes most of the evidence for an AGN. As presented in detail in \cite{Sobral2017}, a re-analysis of old spectra and new near-infrared spectroscopy with the {\it HST}/WFC3 grism shows that the He{\sc ii} line is indeed at lower significance. Moreover, if present, it is at lower luminosity (by a factor $\approx3$) and does not spatially coincide with clump A, but rather is emitted in the direction of clump C. No UV metal lines are detected in clump A. This points towards a moderately low metallicity galaxy that is actively forming stars, without clear evidence for AGN activity (but see further details in \citealt{Sobral2017}).

In this paper, we answer to the question of the presence of metals in CR7, and test whether the metallicity and dust content varies between different UV components. We summarise the UV properties of CR7 in \S $\ref{sec:UVprops}$. ALMA observations, data reduction, astrometry and sensitivity are discussed in \S $\ref{sec:observations}$. We investigate the resolved [C{\sc ii}] emission in \S $\ref{sec:CII}$. Dust continuum measurements and their implications for the SFRs are detailed in \S $\ref{sec:FIR}$. In \S $\ref{sec:SFRLCII}$, we discuss where the different clumps of CR7 are located in the SFR-L$_{\rm [CII]}$ relation, and how this compares with other sources. We discuss the implication of our results in light of recent simulations in \S $\ref{discussion}$, where we also use these observations to update the interpretation of the nature of CR7. The conclusions of this work are summarised in \S $\ref{sec:conclusions}$. Throughout the paper, we assume a $\Lambda$CDM cosmology with $\Omega_M = 0.70$, $\Omega_{\Lambda} = 0.30$ and H$_0 = 70$ km s$^{-1}$ Mpc$^{-1}$, and assume a \cite{Salpeter1995} initial mass function (IMF) with mass limits $0.1$ M$_{\odot}$ and $100$ M$_{\odot}$.

\section{UV properties of CR7}  \label{sec:UVprops}
In the rest-frame UV, CR7 consists of three clumps (A, B and C, \citealt{Sobral2015}), of which the brightest (A) coincides with the peak of Ly$\alpha$ emission, and is also spectroscopically confirmed at $z_{\rm Ly\alpha}=6.604$. UV components are separated by $\sim1''$, corresponding to projected distances of $\sim 5$kpc. The latest photometry on the {\it HST}/WFC3 imaging in the F110W and F160W filters has been performed by \cite{Bowler2016}. We use this photometry to compute UV slopes of the three different clumps individually. The contribution of Ly$\alpha$ to the flux observed in the F110W filter is based on our Ly$\alpha$ narrow-band imaging  as follows: we first correct the Subaru/S-Cam NB921 image for the contribution from the UV continuum by subtracting the $z'$ image (which is calibrated such that a colour of $z'$-NB921=0 corresponds to a line flux of zero, see \citealt{Matthee2015}) and then use this Ly$\alpha$ image to measure the Ly$\alpha$ flux at the positions of clumps A, B and C. The Ly$\alpha$ flux is multiplied by a factor two to take into account that the NB921 filter transmission is 50 \% at the wavelength of CR7's Ly$\alpha$. We then convolve the apertures that have been used for {\it HST} photometry (1$''$, 0.4$''$, 0.4$''$, for clumps A, B and C respectively) with the PSF of the NB921 imaging (0.6$''$) and measure the Ly$\alpha$ flux in those PSF-convolved apertures. We measure Ly$\alpha$ fluxes of $8.3, 2.7, 1.3\times10^{-17}$ erg s$^{-1}$ cm$^{-2}$ in these apertures. Correcting the flux-density observed in the F110W for this line-flux contribution results in corrections of +0.15, +0.16 and +0.05 to the F110W magnitude (and M$_{1500}$) for clumps A, B and C, respectively. We measure UV slopes of $\beta = -2.3\pm0.4, -1.0\pm1.0$ and $-2.3\pm0.8$, for the three clumps respectively (see Table $\ref{tab:measurements}$).\footnote{Without correcting the F110W photometry for the contribution from Ly$\alpha$, we would obtain UV slopes of $\beta = -2.8\pm0.4, -1.5\pm1.0$ and $-2.4\pm0.8$ for clump A, B and C. While the errors are still dominated by measurement uncertainty, systematic uncertainties due to ignoring potentially important contributions from Ly$\alpha$ could bias UV slope measurements to artificially bluer values.} Due to the shallow depth of the observations, the UV slopes of clumps B and C are only poorly constrained, and deeper observations particularly in the F160W filter are required to improve them. 
\begin{figure}
\includegraphics[width=8.6cm]{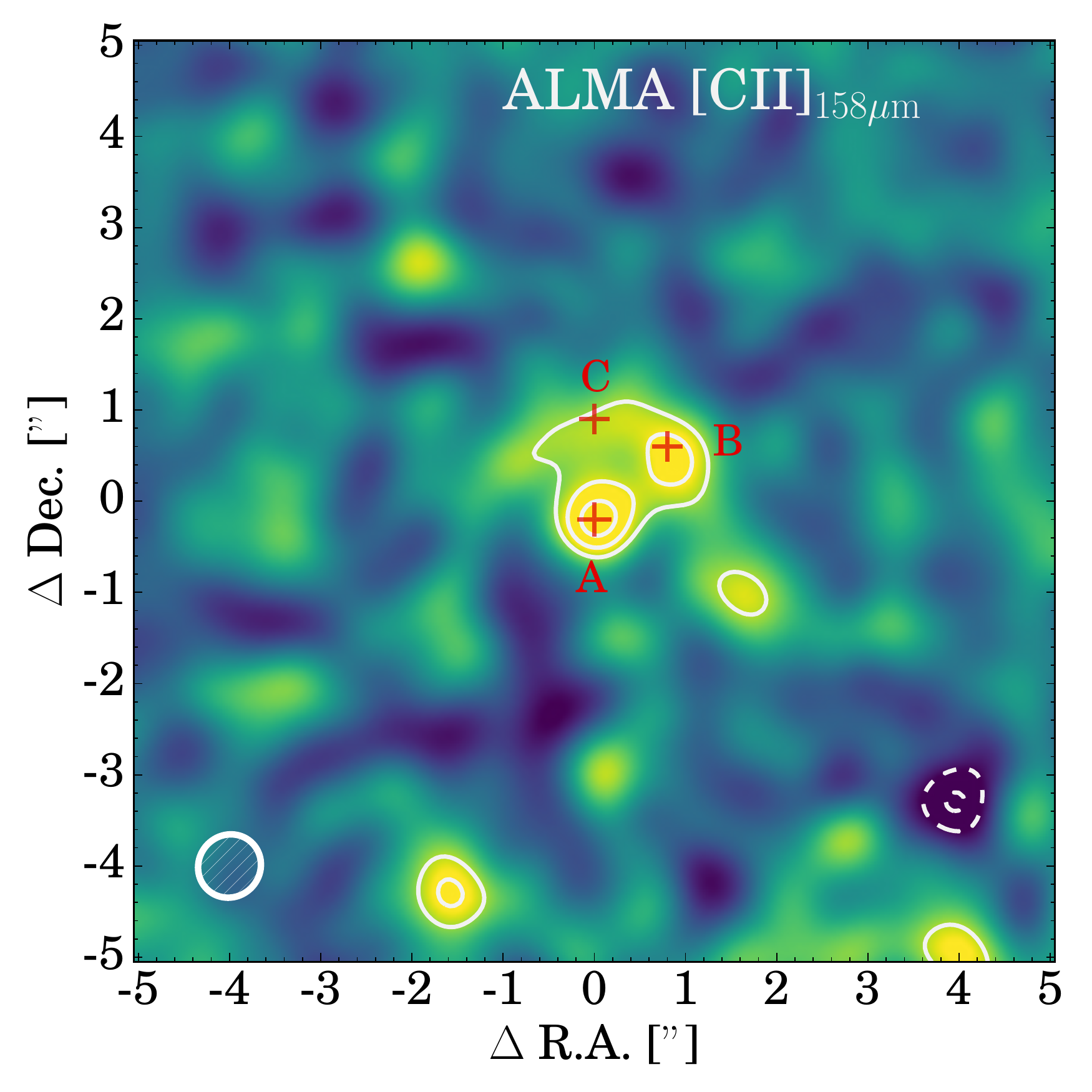}
\caption{ALMA [C{\sc ii}] narrow-band image of CR7, collapsed for frequencies with $-614$ to $+96$ km s$^{-1}$ with respect to the Ly$\alpha$ reference frame ($z=6.604$). The image ($54\times54$ kpc) is centred on the peak of Ly$\alpha$ position and the ALMA astrometry is shifted by $0.25''$ to the west to align it with {\it HST} imaging as described in \S $\ref{sec:astrometry}$. Contours are shown at the $3, 4, 5 \sigma$ significance levels (with 1$\sigma = 0.016$ Jy beam$^{-1}$ km s$^{-1}$). Negative contours are shown as dashed lines. The beam size is shown in the bottom left corner and has FWHM $0.82''\times0.77''$. The red crosses mark the positions of the UV clumps of CR7. [C{\sc ii}] is clearly detected in CR7, indicating the presence of carbon.  \label{fig:ALMA}}
\end{figure}

Based on the rest-frame UV luminosities and UV slopes, we derive SFR$_{\rm UV}$, where dust attenuation is estimated using the \cite{Meurer1999} attenuation law. We take the measurement uncertainties in the rest-UV magnitudes into account by perturbing these values 10,000 times (assuming the uncertainties are gaussian). In each realisation, we re-compute the UV luminosity and UV slope, and use these to derive dust-corrected SFR. We then obtain the median and the 1$\sigma$ percentiles and use these to derive the asymmetric uncertainties.
With this method, the SFRs are $29^{+23}_{-2}$, $38^{+182}_{-32}$ and $7^{+19}_{-1}$ M$_{\odot}$ yr$^{-1}$ for clumps A, B and C, respectively. The uncertainties in the dust corrected SFRs are large due to the propagation of errors in $\beta$. However, as we show in \S $\ref{sec:FIR}$, these uncertainties are mitigated by constraints on the IR luminosity from our deep ALMA observations, which place firm limits on the dust-obscured SFR. Because of these constraints, our final results would also only change marginally if an SMC-like attenuation law is used. The results are listed in Table $\ref{tab:measurements}$. The F110W imaging has a 3$\sigma$ sensitivity of 27.3 AB magnitude in a 0.4$''$ diameter, which corresponds to a limiting UV magnitude M$_{1500} > -19.5$ and a SFR$_{\rm UV} < 4$ M$_{\odot}$ yr$^{-1}$.

\section{ALMA data} \label{sec:observations}
\subsection{Observations \& data reduction} 
We observed CR7 (10:00:58.00 +01:48:15.3, J2000) in band 6 with ALMA during cycle 3 with configuration C43-4, aimed to achieve a 0.3$''$ angular resolution (program ID  \#2015.1.00122.S). The target has been observed with between 35-43 antennas for a total on source integration time of 6.0h on 22, 23, 24 May and 4, 8 November 2016, with precipitable water vapor ranging from 0.3-1.4 mm. We observed in four spectral windows centred at 249.94592, 247.94636, 234.94917 and 232.9496 GHz with a bandwidth of 1875 MHz, of which the first is centred at [C{\sc ii}] line emission ($\nu_{z=6.604} = 249.9395$ GHz). The central frequencies of the spectral windows correspond to rest-frame frequencies between 1771.7 GHz and 1901.0 GHz, with a velocity resolution of 19.4 km s$^{-1}$. The phase calibrations have been performed on the source J0948+0022. The quasar J1058+0133 has been observed regularly as a bandpass and flux calibrator, resulting in typical flux calibration errors of $\sim10$ \%. Data have been reduced using {\sc Casa} version 4.7.0 \citep{McMullin2007} with natural weighting and channel averaging resulting in a 38.8 km s$^{-1}$ velocity resolution. We use uv tapering of the visibilities with a Gaussian width FWHM=0.7$''$ to optimise the S/N ratio. The observational set-up and data reduction results in a beam size of $0.35\times0.33''$ (gaussian $\sigma$; $0.82\times0.77''$ FWHM), with a position angle of $-49.5 \degree$. We have also performed the reduction with uv tapering with different smoothing kernels ranging from $0.4-0.8''$, but find that all changes in luminosity, line-width and size measurements are consistent within the error-bars. A simple estimate of the noise level of the pixels within a radius of 20$''$ of the center results in rms=0.06 mJy beam$^{-1}$ in 38.8 km s$^{-1}$ channels.

\begin{figure}
\includegraphics[width=8.7cm]{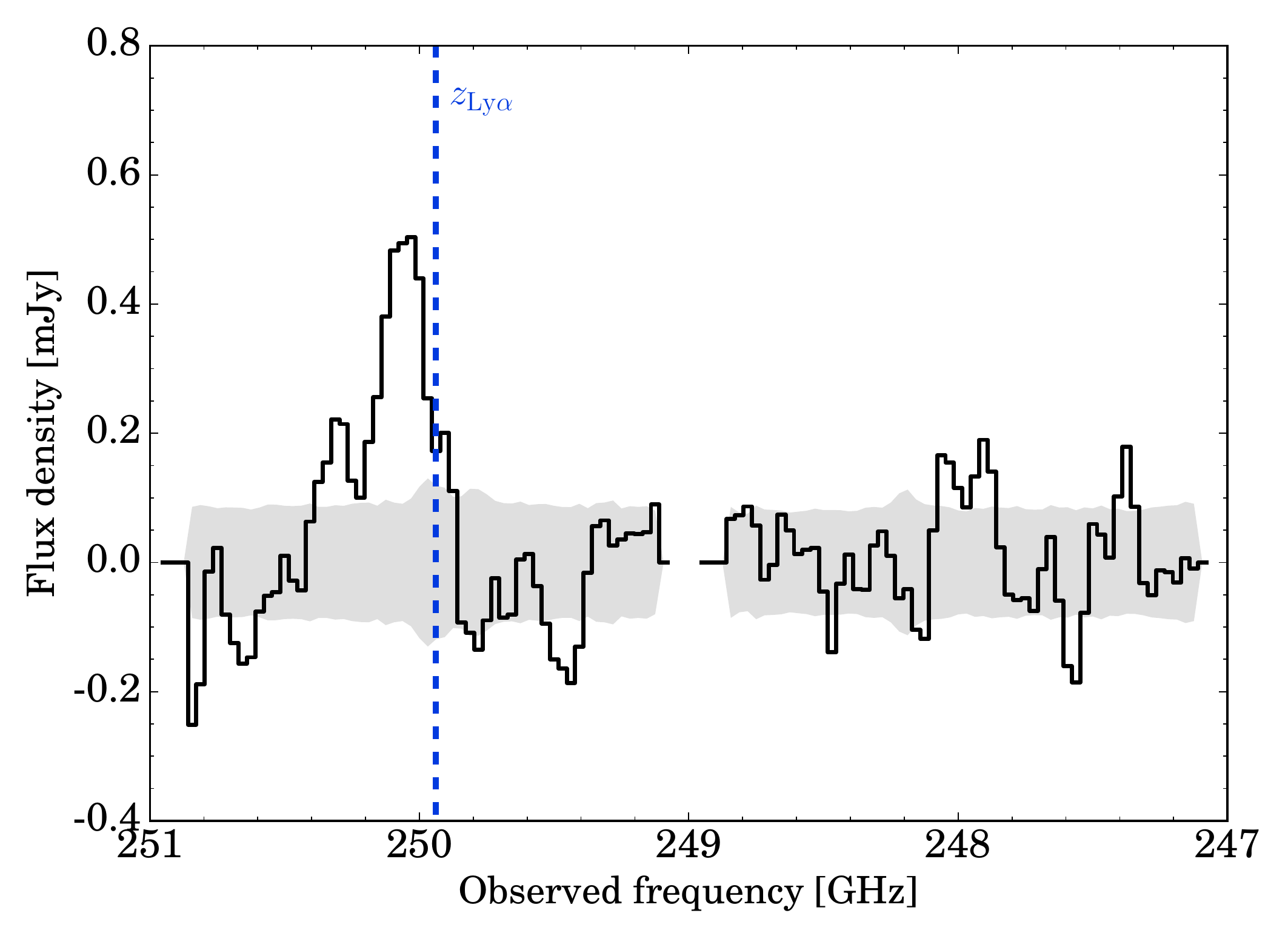}
\caption{Integrated spectrum over the area within the 3$\sigma$ [C{\sc ii}] contours in Fig. $\ref{fig:ALMA}$, in two observed spectral windows. The grey shaded region indicates the 1$\sigma$ error estimated using apertures with the same geometry as the 3$\sigma$ contour area, as a function of frequency. The blue dashed line indicates where [C{\sc ii}] would be detected at $z=6.604$, the Ly$\alpha$ redshift of CR7. Ly$\alpha$ is redshifted by $\approx -170$ km s$^{-1}$ with respect to the systemic [C{\sc ii}] redshift.  \label{fig:CII_spectrum}}
\end{figure}

\subsection{Astrometry} \label{sec:astrometry}
As described in \S $\ref{sec:CII}$, we detect two separate clumps at high significance in the [C{\sc ii}] image collapsed over all velocities at which significant line-emission is detected, see Fig. $\ref{fig:ALMA}$. These [C{\sc ii}] clumps are offset by 0.25$''$ to the east with respect to the UV positions of clumps A and B, while the offset in declination is $<0.10''$, respectively. The alignment of both clumps, and their separation is also similar in the UV and in [C{\sc ii}]. As these two [C{\sc ii}] clumps resemble the geometry of clumps A and B, we assume that the offsets between the UV positions and [C{\sc ii}] positions are due to errors in the astrometry. This offset is similar in magnitude as reported in e.g. \cite{Dunlop2017}, but in the R.A. direction instead of the Dec. direction (see also \citealt{Carniani2017} and references therein for offsets in $z\sim6-7$ galaxies). For the rest of the analysis, we align the ALMA data with {\it HST} and Subaru narrow-band data by applying an offset of 0.25$''$ to the west. We note that the positions of two serendipitous IR continuum-detections of foreground galaxies are also in agreement with this offset (see \S $\ref{sec:FIR_blind}$ and Fig. $\ref{fig:continuum_map}$).
\begin{figure}
	\includegraphics[width=8.7cm]{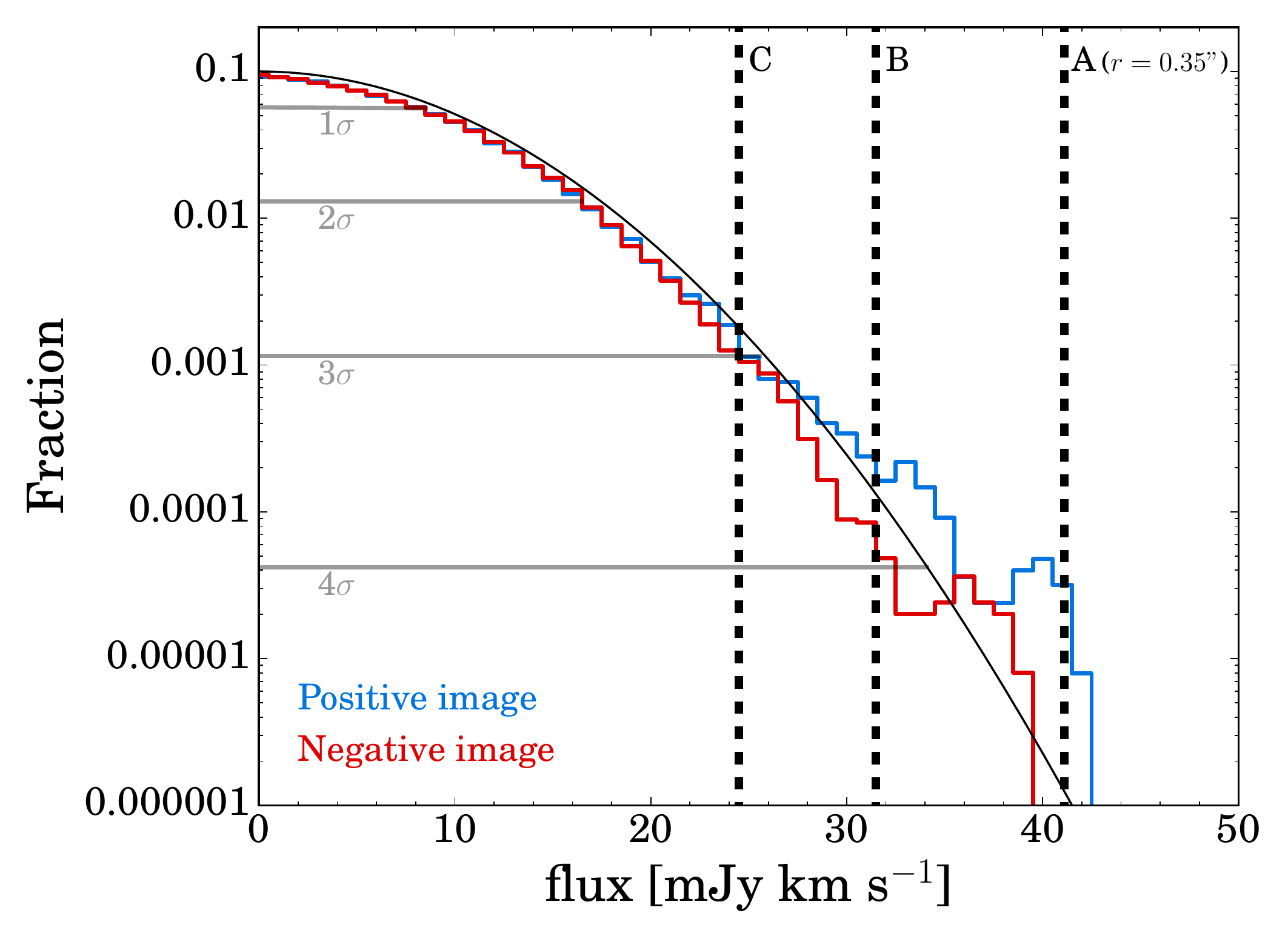}
\caption{Histogram of the integrated [C{\sc ii}] fluxes measured over circular, beam-sized scales at random positions in the collapsed cube (with velocities from $-614$ to $+96$ km s$^{-1}$ with respect to Ly$\alpha$). The blue histogram shows positive fluxes, while the red histogram shows negative fluxes. We indicate the [C{\sc ii}] flux measured at the positions of {\it HST} clumps A, B and C with vertical lines. These correspond to $\approx 5, 4, 3$ $\sigma$ detections, respectively. The grey curve shows the expected noise properties if the noise is gaussian with a 1$\sigma$ dispersion. Note that our final measurement of clump A uses a larger aperture (with radius 0.55$''$), resulting in a similar significance but higher flux. Also note that the detections are at slightly higher significance in narrower-velocity slices optimised for each clumps specifically.  \label{fig:significance}}
\end{figure}

\begin{figure*}
\centering
\begin{tabular}{ccc}
	\includegraphics[width=5.7cm]{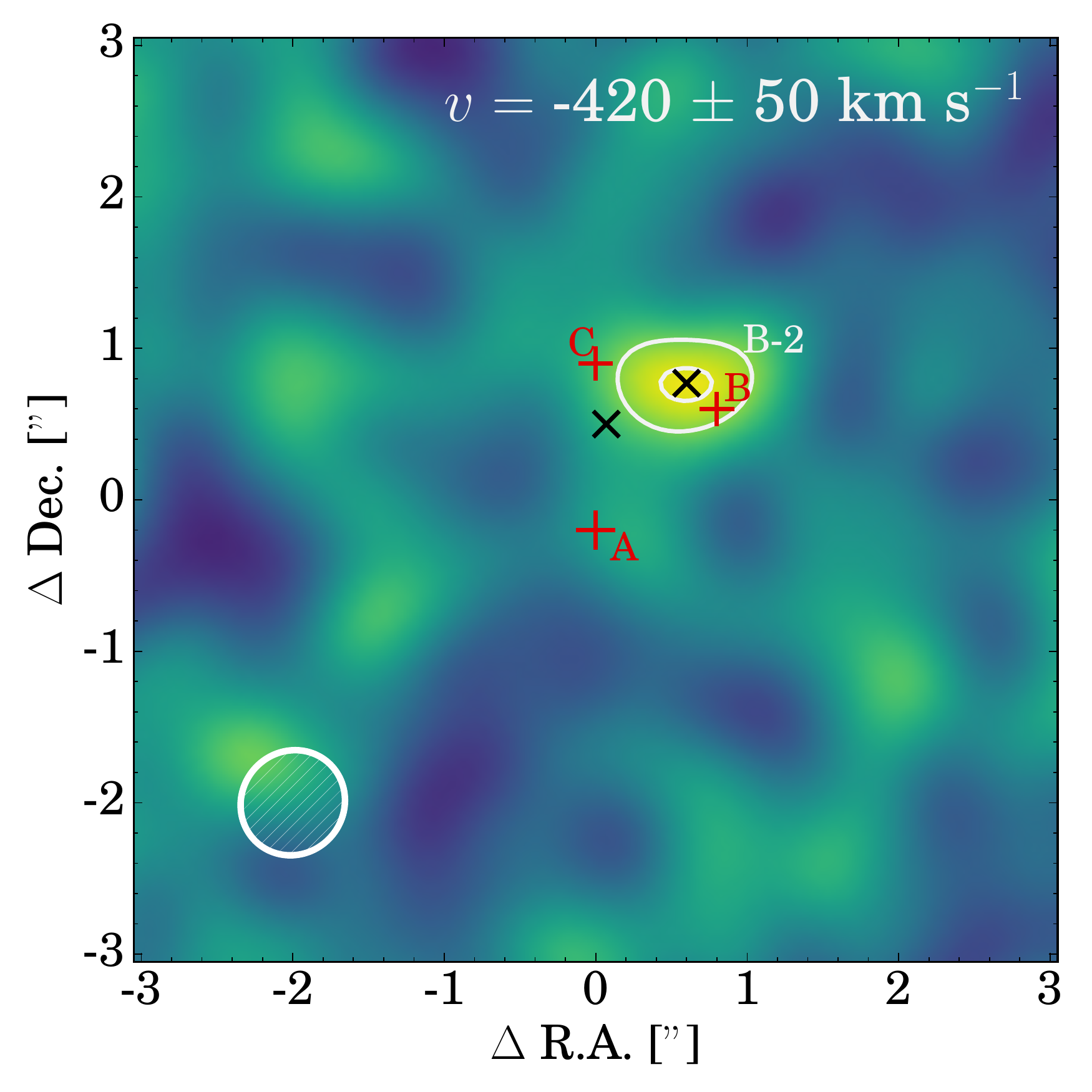}&
	\includegraphics[width=5.7cm]{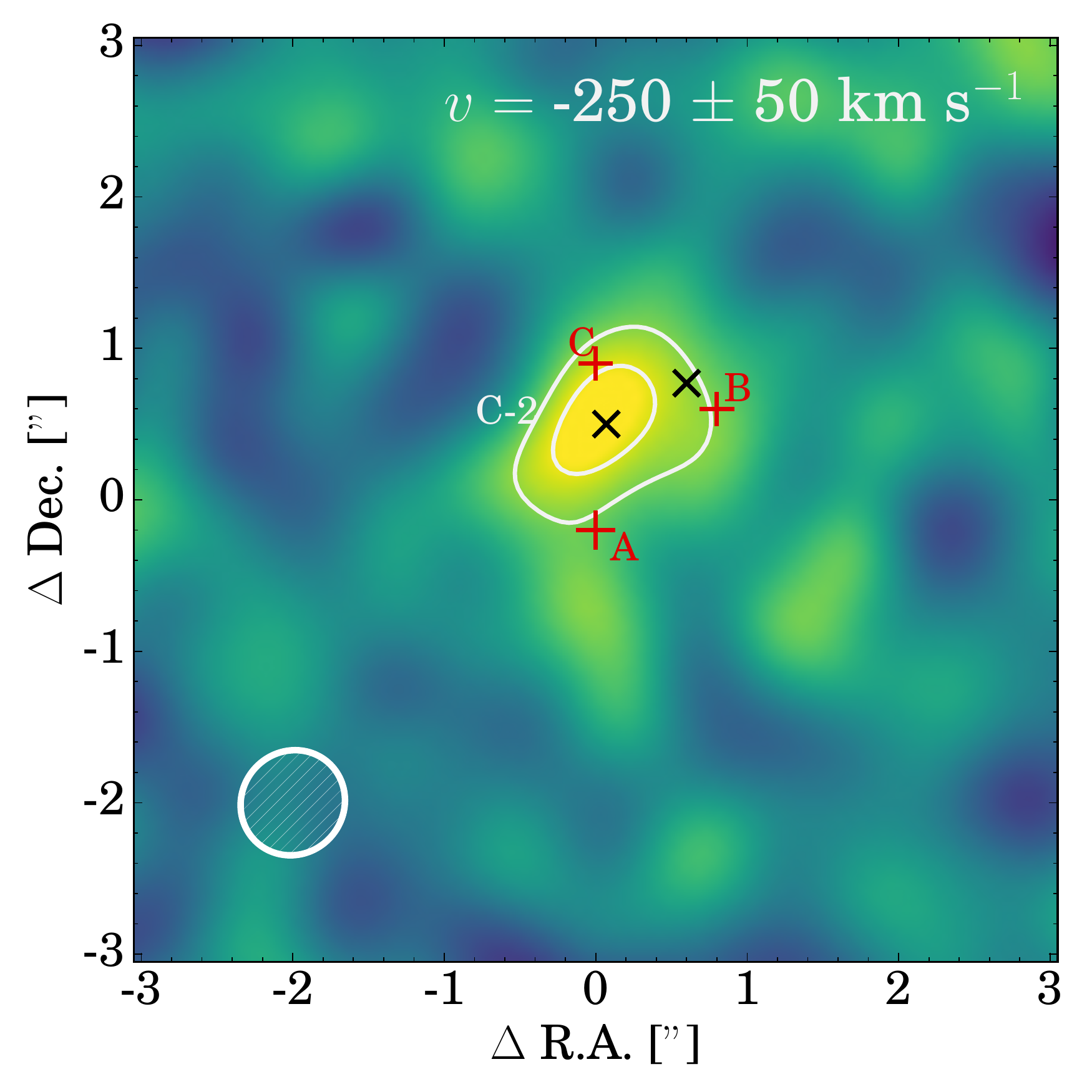}&
	\includegraphics[width=5.7cm]{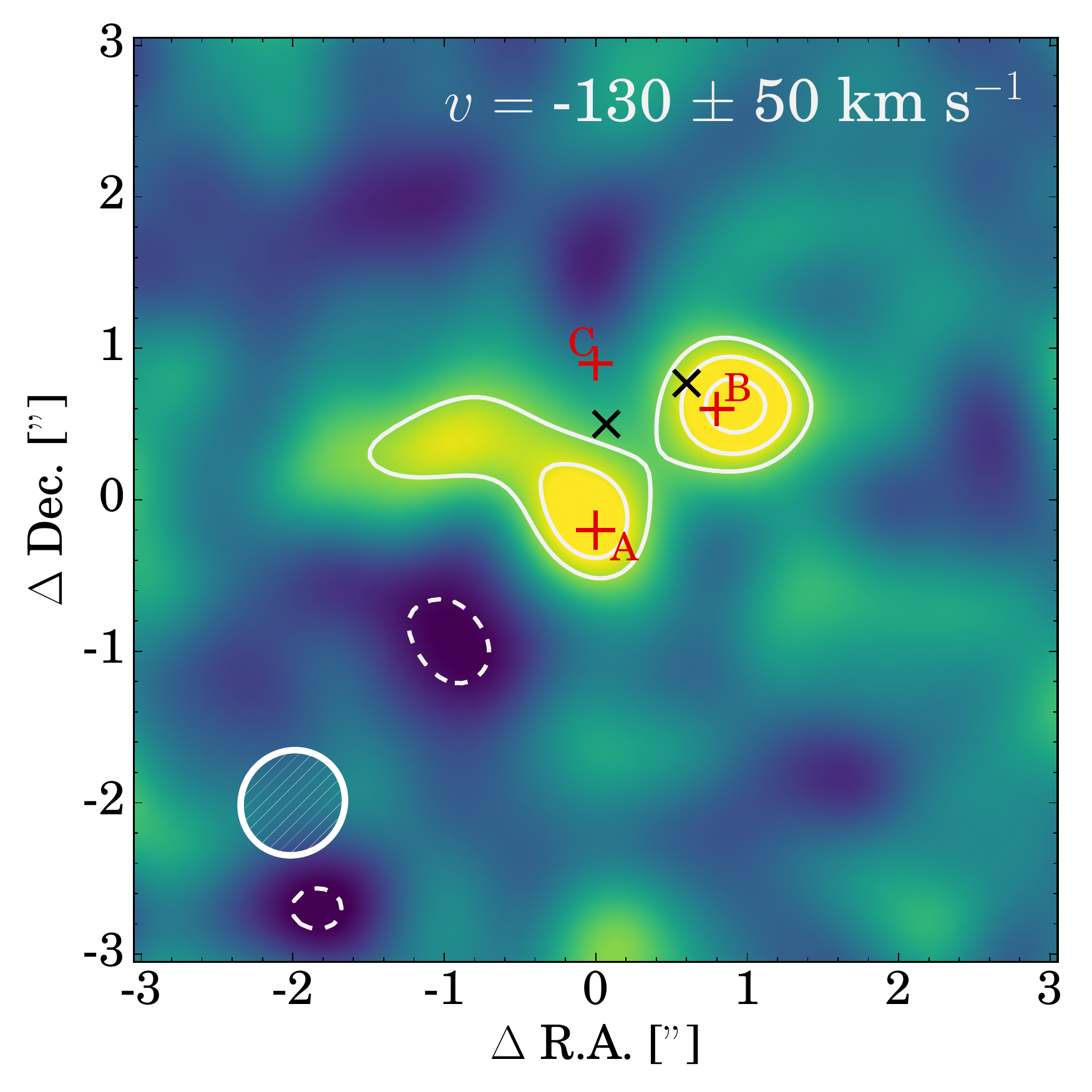}\\
	\end{tabular}
    \caption{[C{\sc ii}] map in three velocity channels of 100 km s$^{-1}$, zoomed in on CR7 (with a scale of $32\times32$ kpc) and with velocities with respect to the Ly$\alpha$ redshift of $z=6.604$. Contours mark the 3, 4, 5 $\sigma$ levels, with sensitivities $\sigma\approx5-6$mJy beam$^{-1}$ km s$^{-1}$. Red crosses mark the positions of the UV clumps, while black crosses mark the positions of additional [C{\sc ii}] components B-2 and C-2. The most luminous [C{\sc ii}] components coincide with UV clumps A and B and have a similar peak velocity. Fainter [C{\sc ii}] flux is detected at $>4\sigma$ in between clumps A, B and C offset by $-120$ km s$^{-1}$  (clump C-2), and close to clump B at $-290$ km s$^{-1}$ (clump B-2).}
    \label{fig:velocity_2ds}
\end{figure*}

\subsection{Line-sensitivity measurements} \label{error_estimate}
As the noise is correlated spatially and spectrally, we measure the sensitivity of the observations and significance of detections as follows. In each collapsed image (which has a size of $51.2''\times51.2''$, centred on CR7), we measure the noise over the same spatial scales as those used for measurements (typically a beam-size, $r=0.35''$). We place circular apertures on 100,000 random positions on the image and integrate the flux over the scales confined by these apertures. As the image size is much larger than the source size, we do not mask any central region. We then compute the 1$\sigma$ detection significance using the r.m.s., which results in 16.9 mJy beam$^{-1}$ km s$^{-1}$ for beam-sized apertures on the collapsed image shown in Fig. $\ref{fig:ALMA}$. 

As illustrated in Fig. $\ref{fig:significance}$, the distribution of fluxes measured on random positions approximately follows a gaussian. It can be seen that the negative image (where we have inverted all counts) has slightly lower number counts than the positive image at high flux levels. This is due to random positions around the centre of CR7, or real sources around CR7. These sources (such as a $\sim 4\sigma$ detection 4$''$ from the south-east of CR7, see Fig. $\ref{fig:ALMA}$) are discussed and investigated in a future paper. When integrating over a larger spatial scale, the noise estimate is slightly higher due to spatially correlated noise. The noise level is lower when data is collapsed over a smaller velocity range. In individual velocity channels (with resolution 38.8 km s$^{-1}$), we measure a 1$\sigma$ sensitivity between 30-40 $\mu$Jy when integrating flux over beam-sized scales.
Throughout the paper, we convert measured fluxes to [C{\sc ii}] luminosities following e.g. \cite{Solomon1992} and \cite{CarilliWalter2013}:
\begin{equation}
{\rm L}_{\rm [CII]}/{\rm L}_{\odot} = 1.04 \times 10^{-3} \times \rm S_{\nu} \Delta v \times \nu_{\rm obs} \times D^2_{L} ,
\end{equation}
where $S_{\nu} \Delta \rm v$ is the flux in Jy km s$^{-1}$, $\nu_{\rm obs}$ the observed frequency in GHz ($\approx 250.0$ GHz for CR7) and $\rm D_L$ is the luminosity distance in Mpc (64457.8 Mpc at $z=6.60$).

\section{Resolved [C{\sc ii}] emission} \label{sec:CII}
\subsection{Total luminosity}\label{total_cii}
\begin{figure*}
\centering
\begin{tabular}{cccc}
	\includegraphics[width=4.2cm]{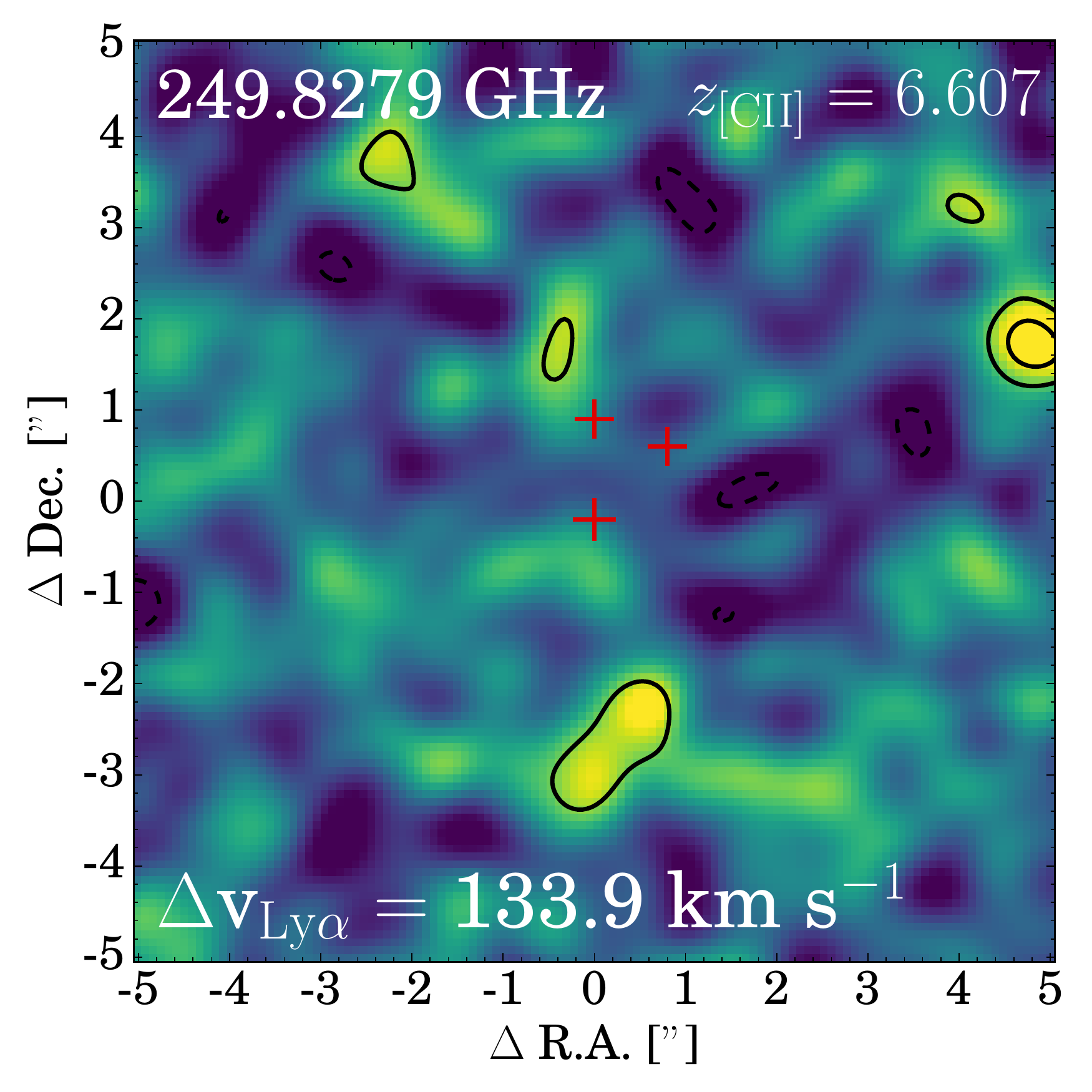}&
	\includegraphics[width=4.2cm]{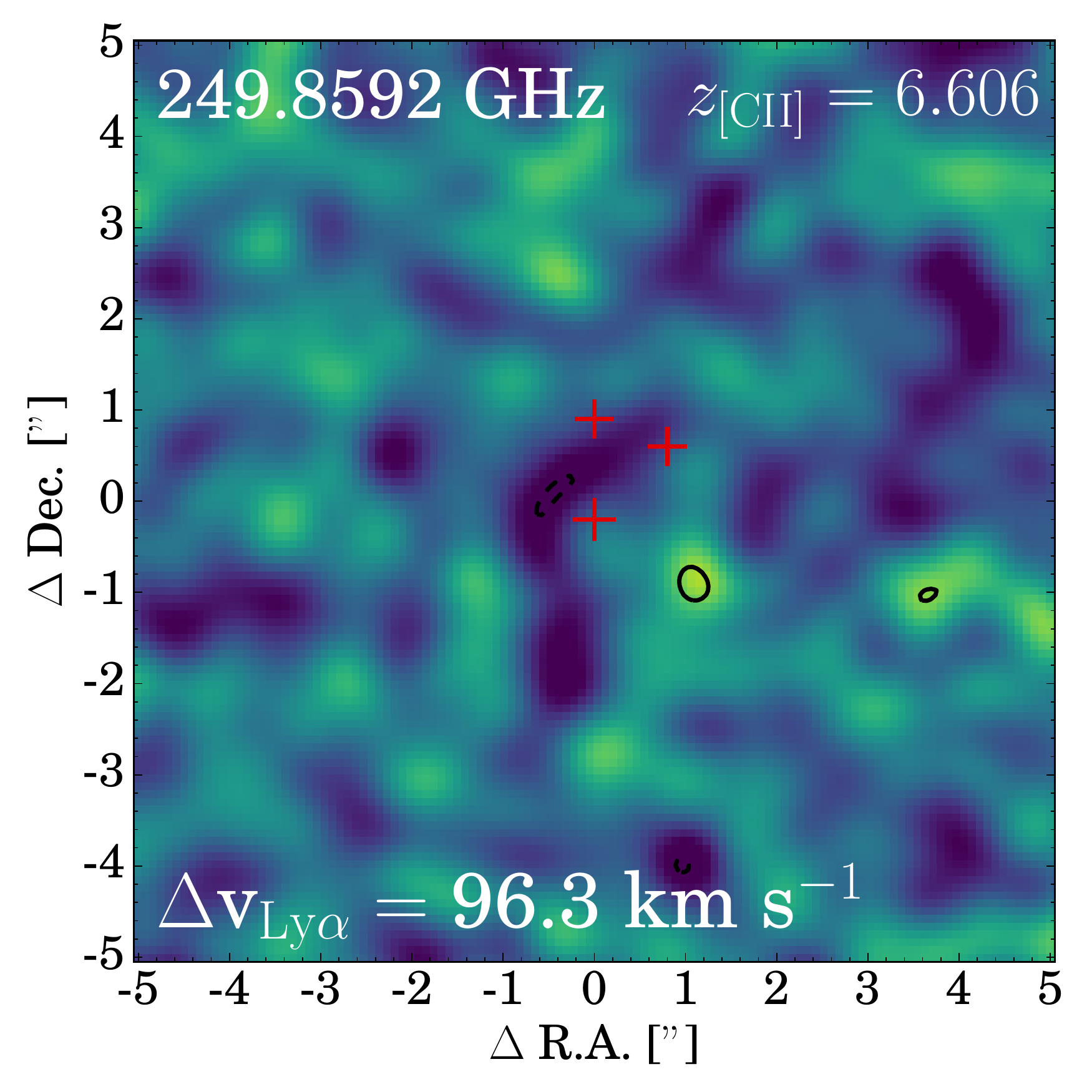}&
	\includegraphics[width=4.2cm]{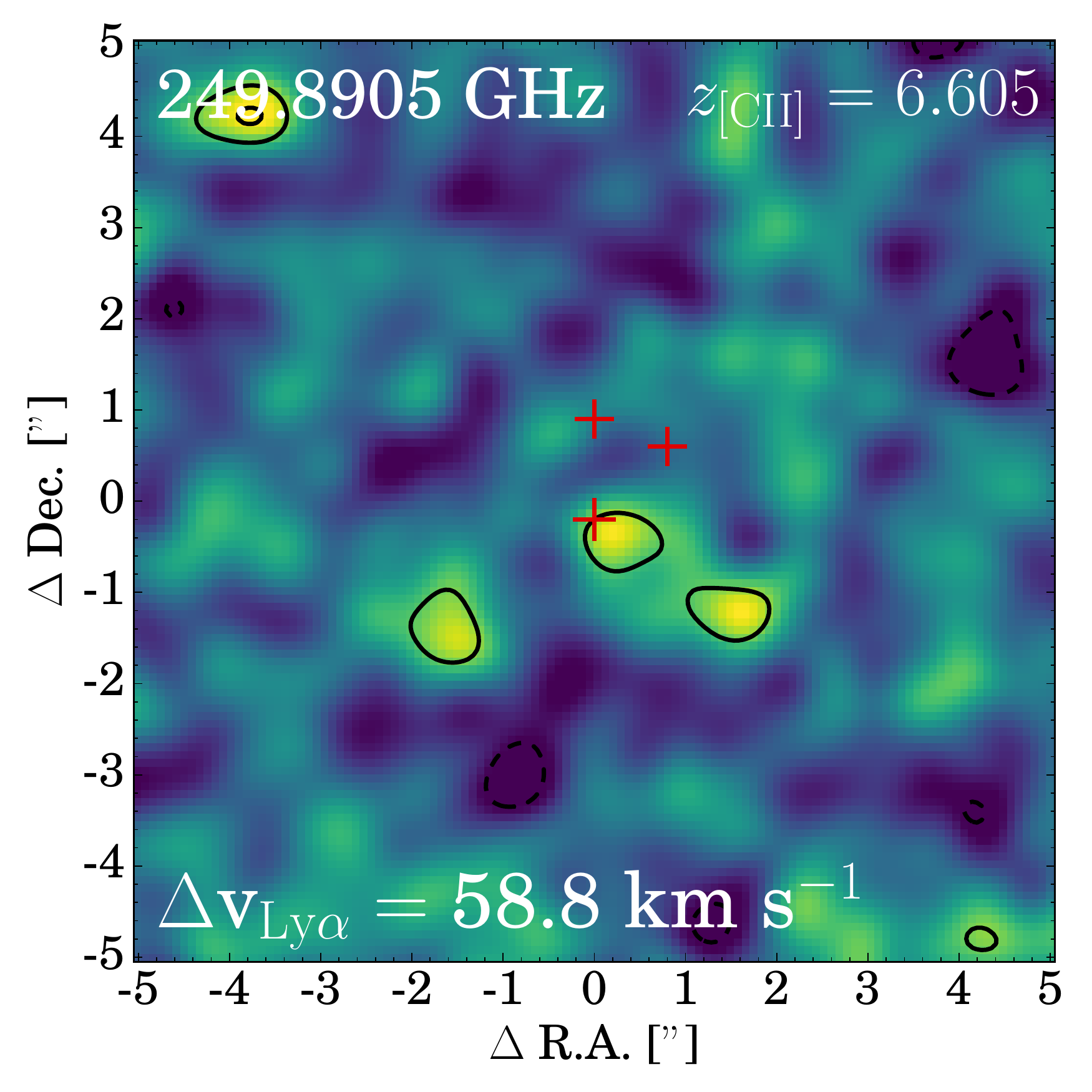}&
	\includegraphics[width=4.2cm]{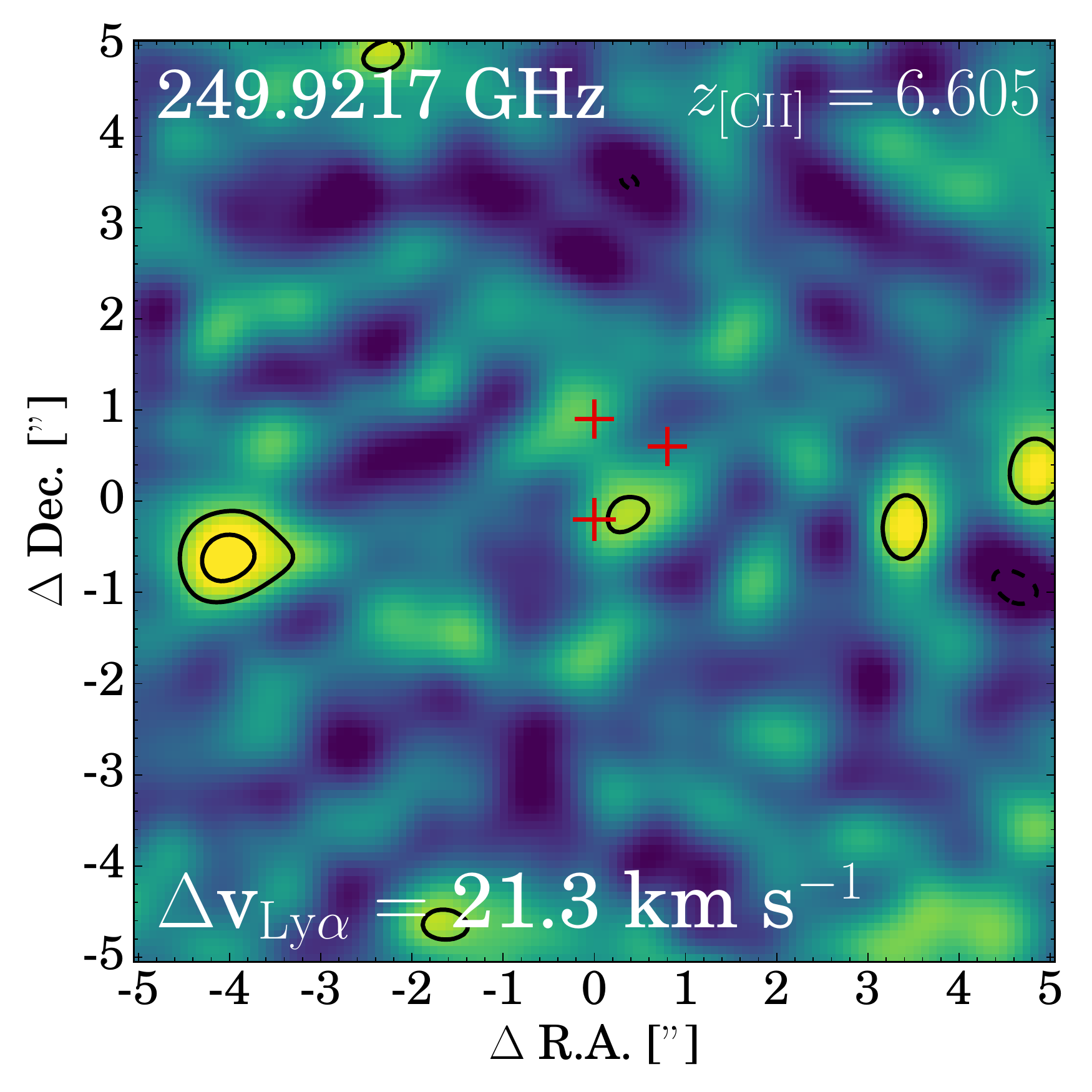}\\
	\includegraphics[width=4.2cm]{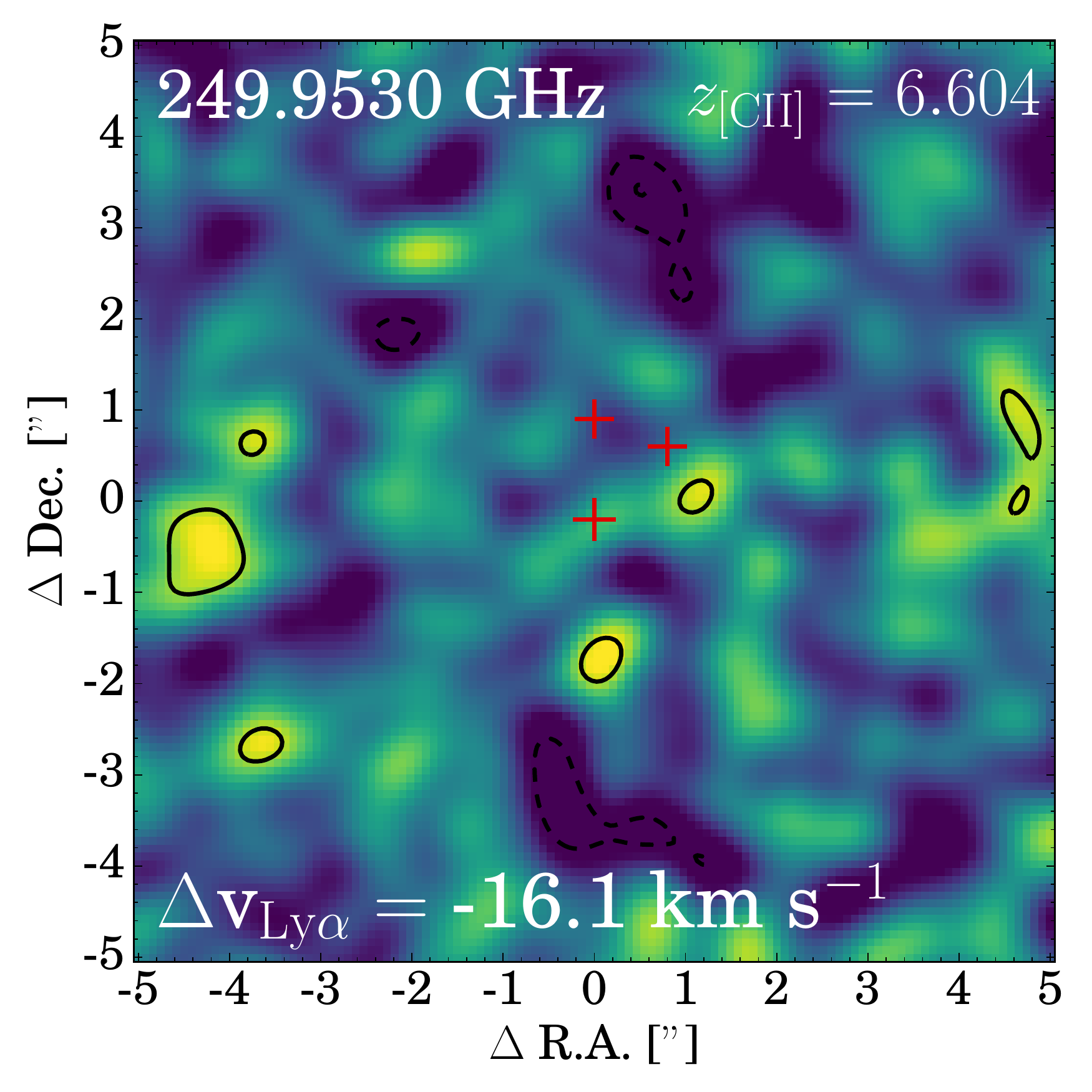}&
	\includegraphics[width=4.2cm]{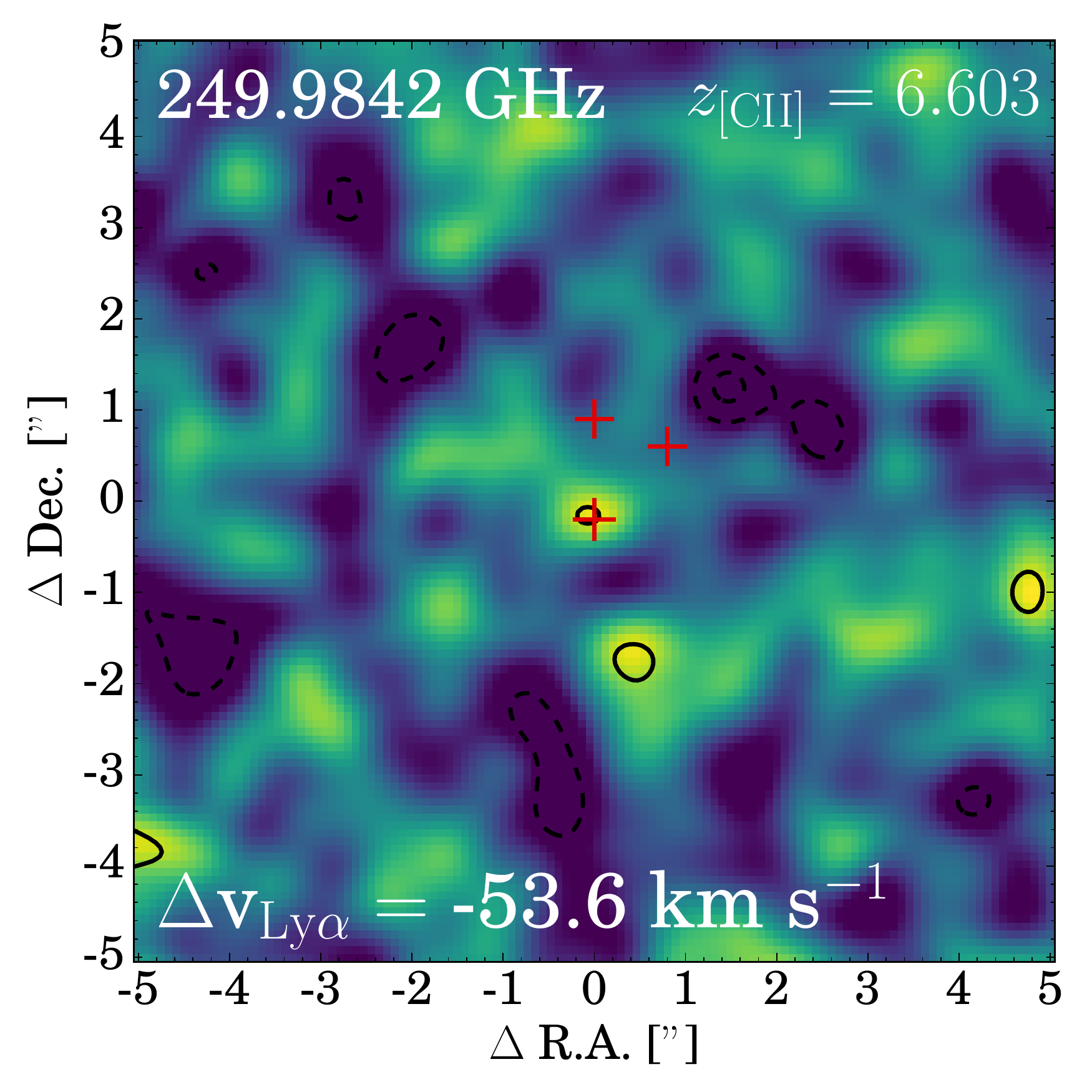}&
	\includegraphics[width=4.2cm]{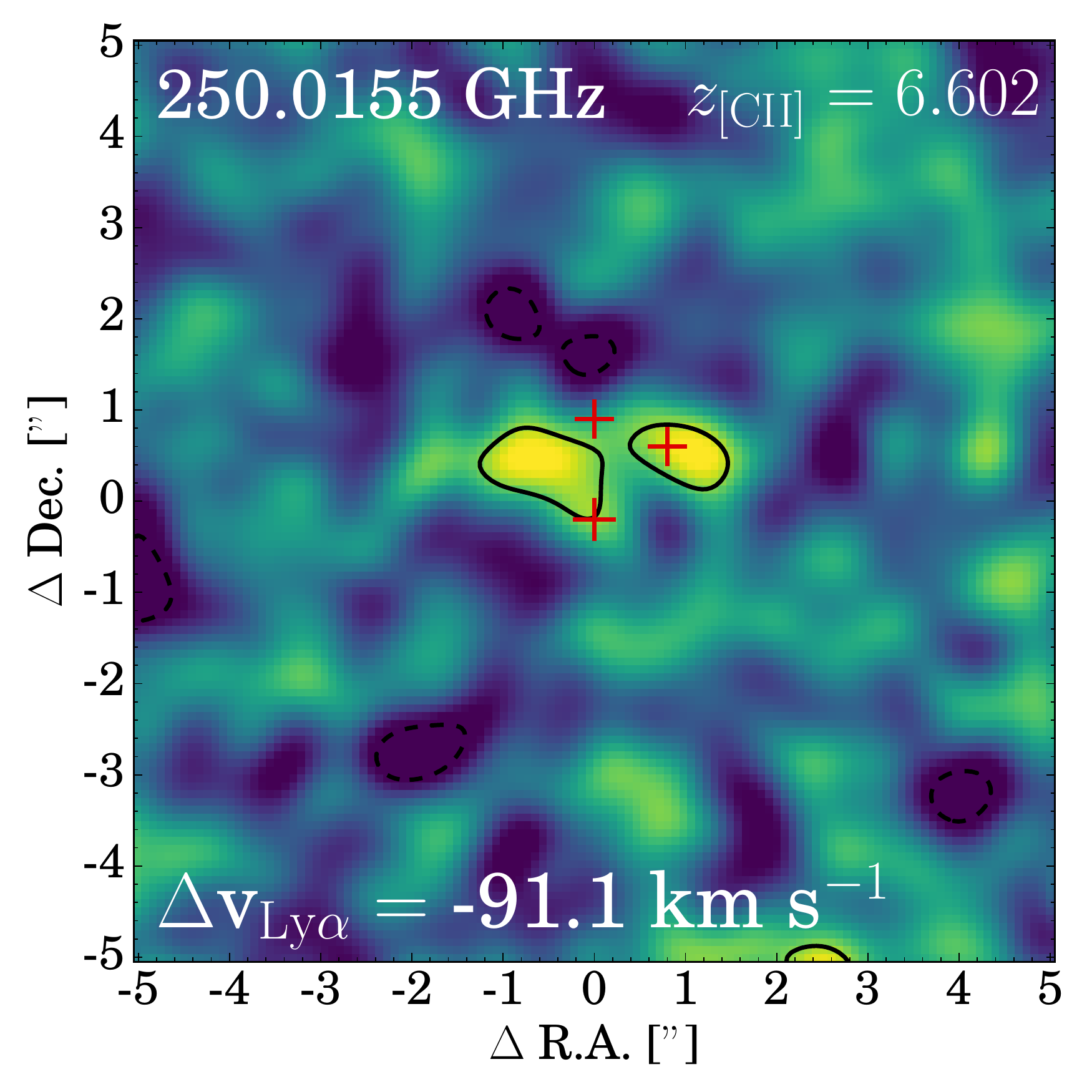}&
	\includegraphics[width=4.2cm]{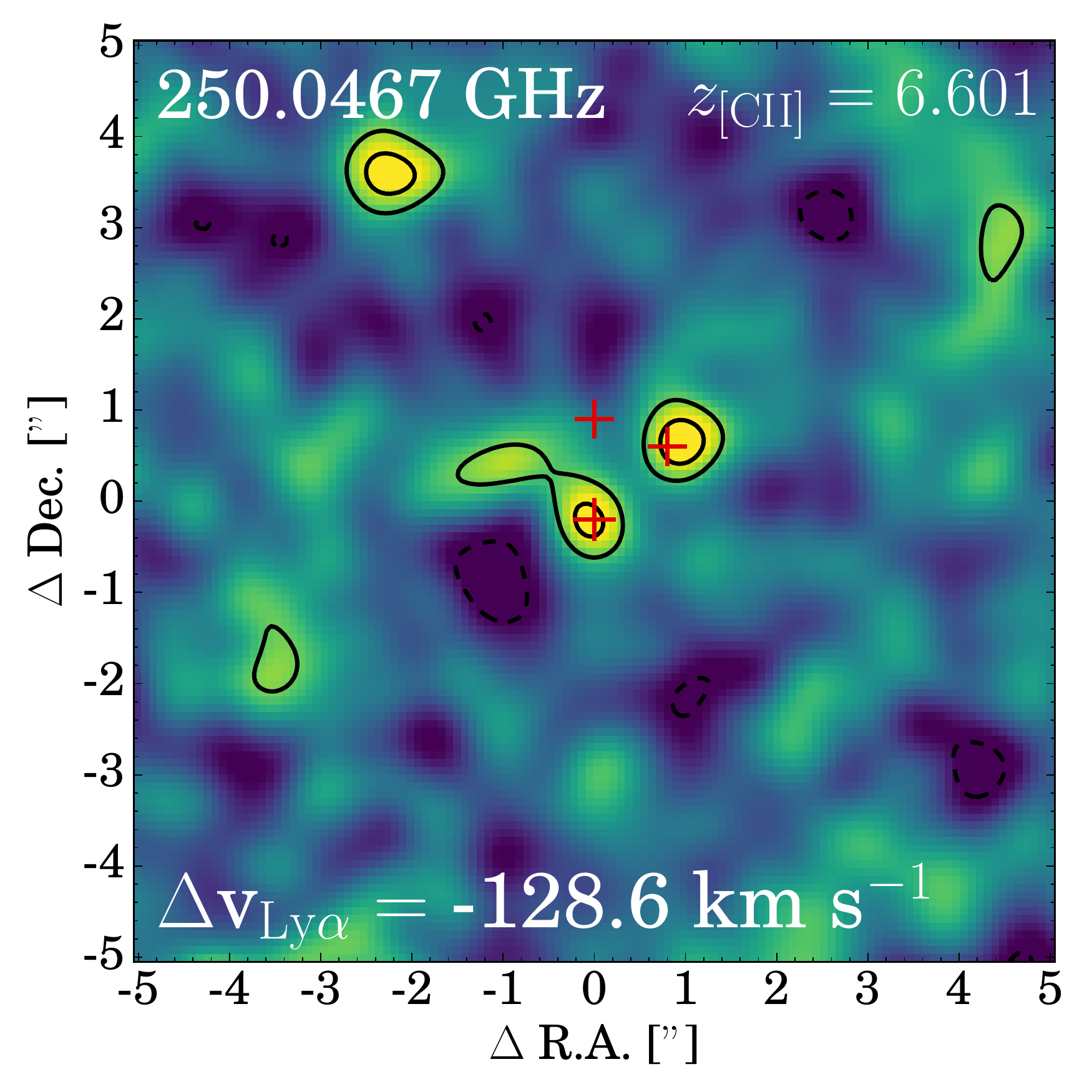}\\
	\includegraphics[width=4.2cm]{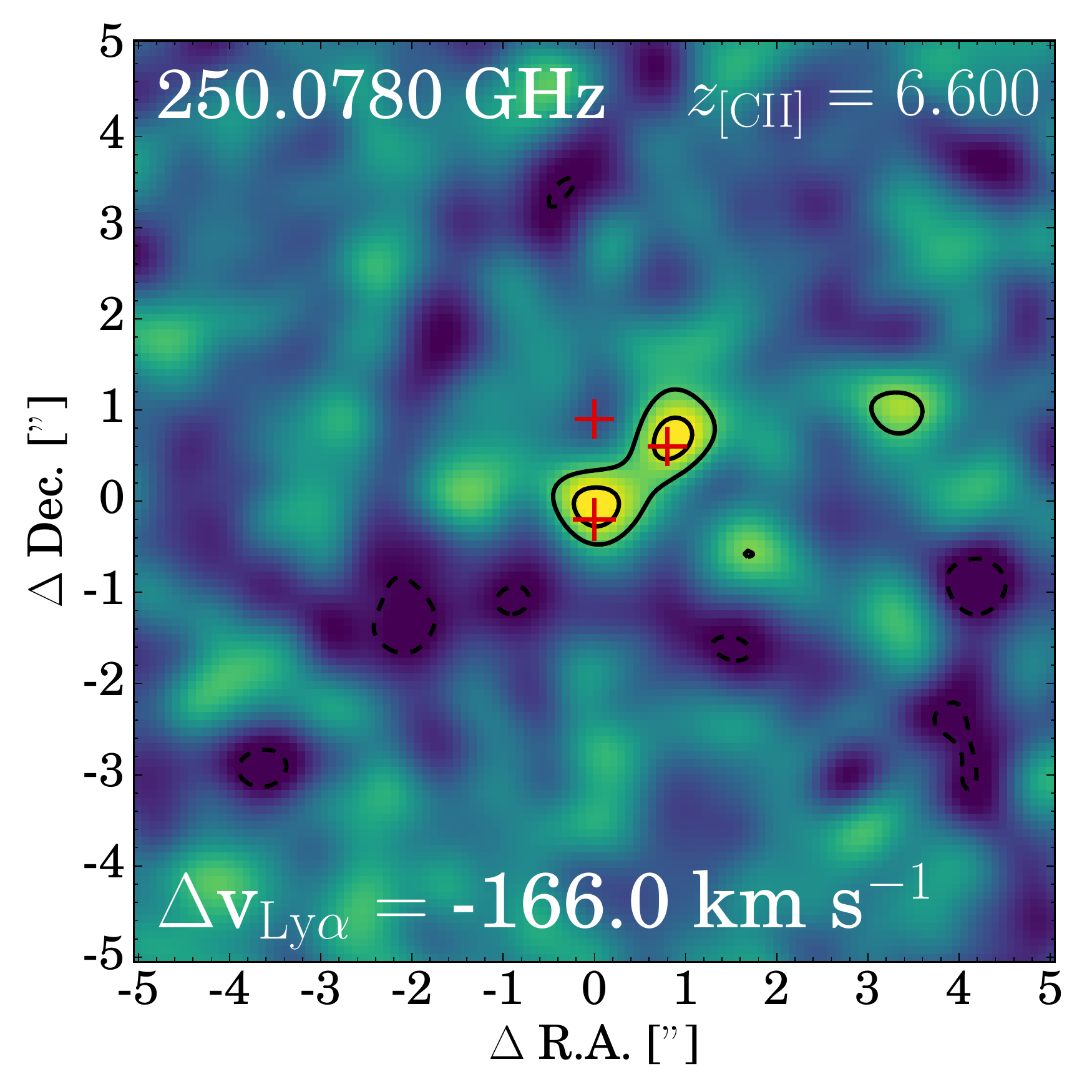}&
	\includegraphics[width=4.2cm]{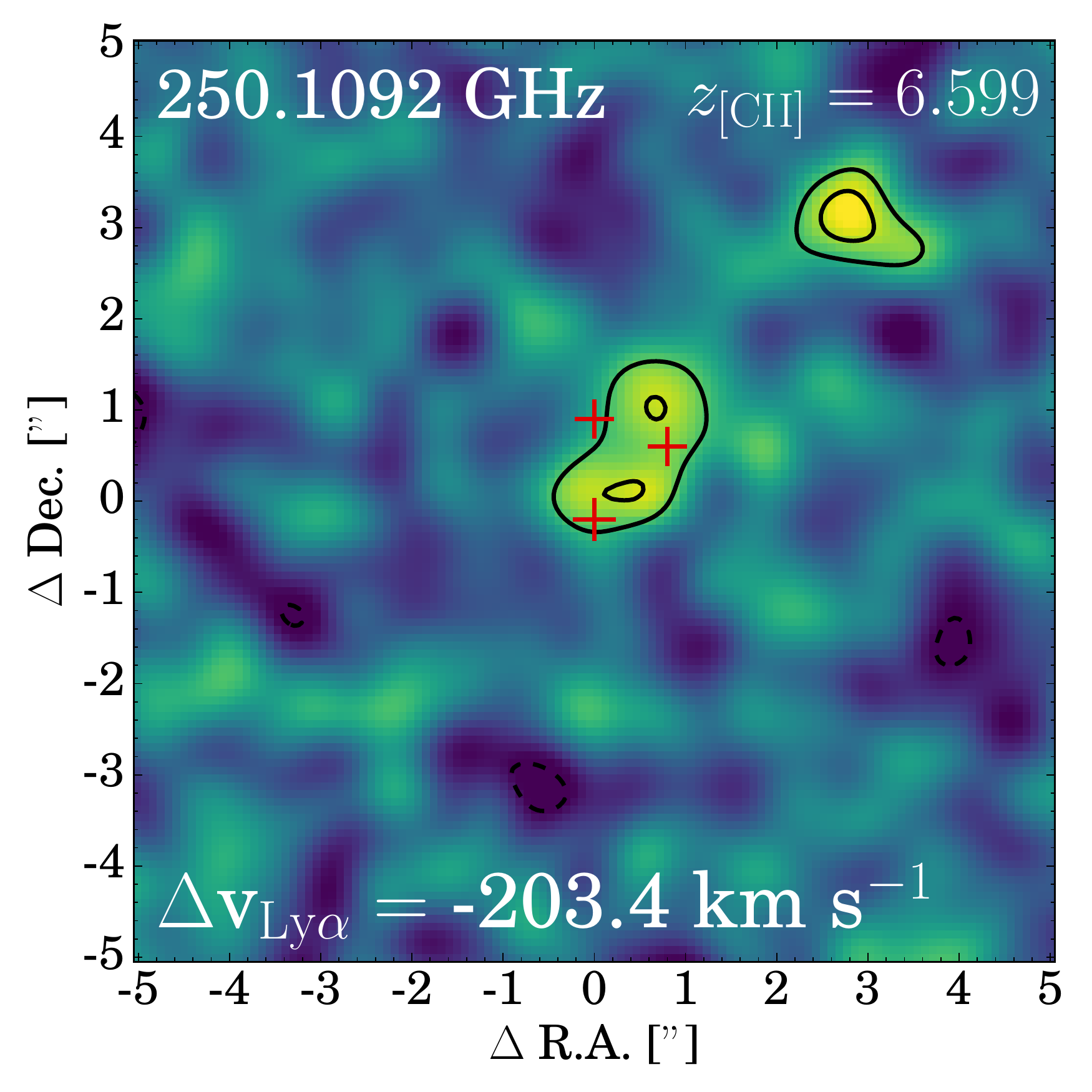}&
	\includegraphics[width=4.2cm]{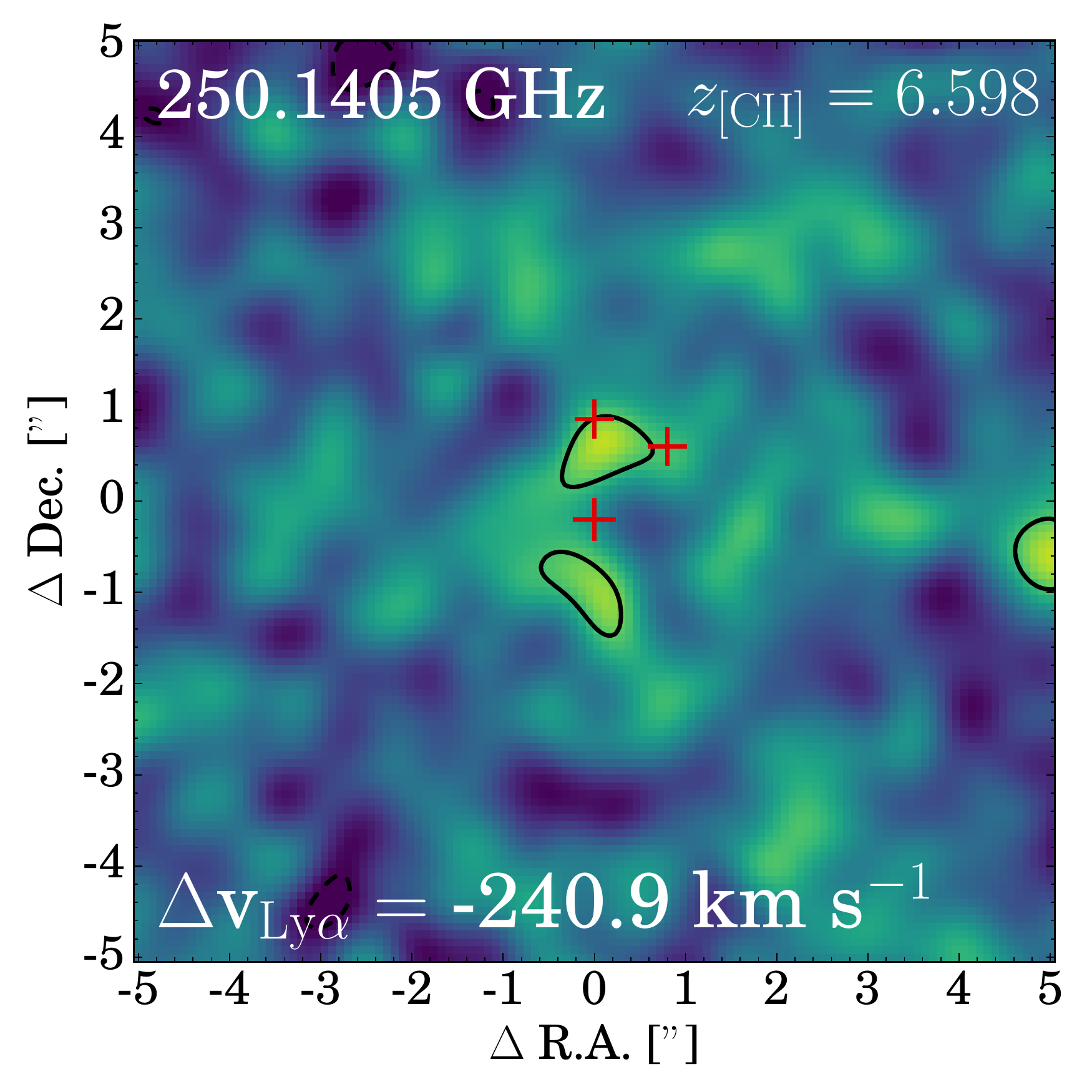}&
	\includegraphics[width=4.2cm]{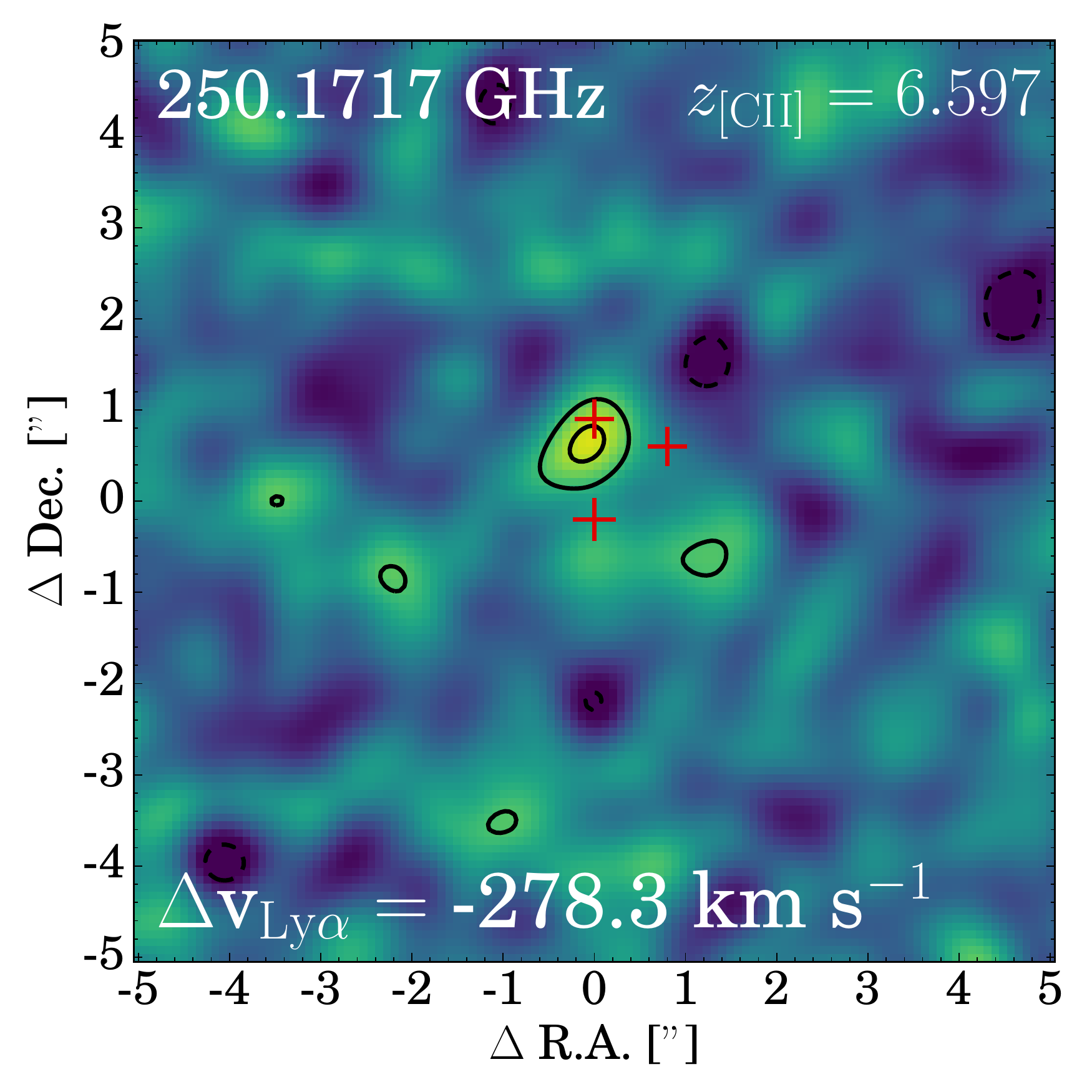}\\
	\includegraphics[width=4.2cm]{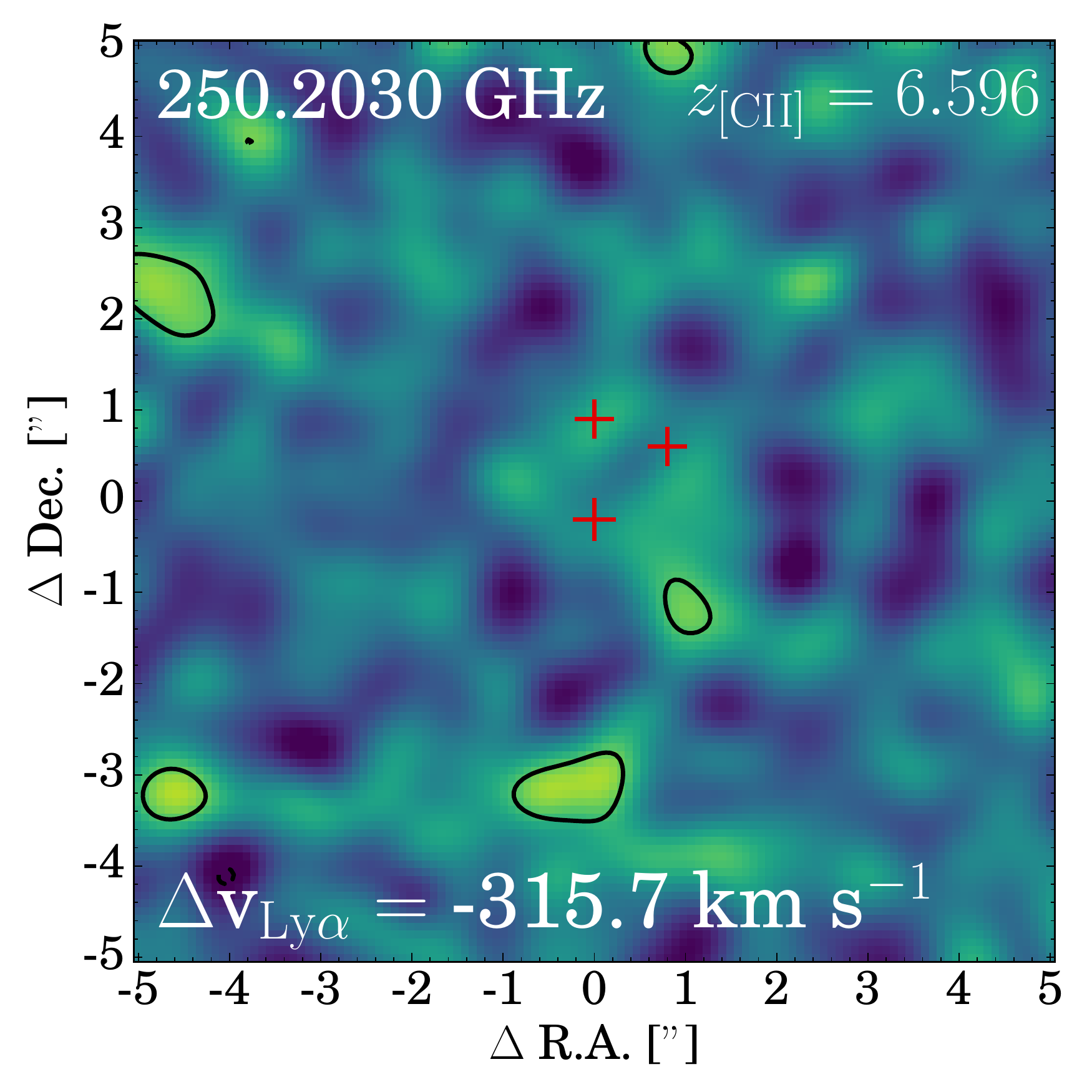}&
	\includegraphics[width=4.2cm]{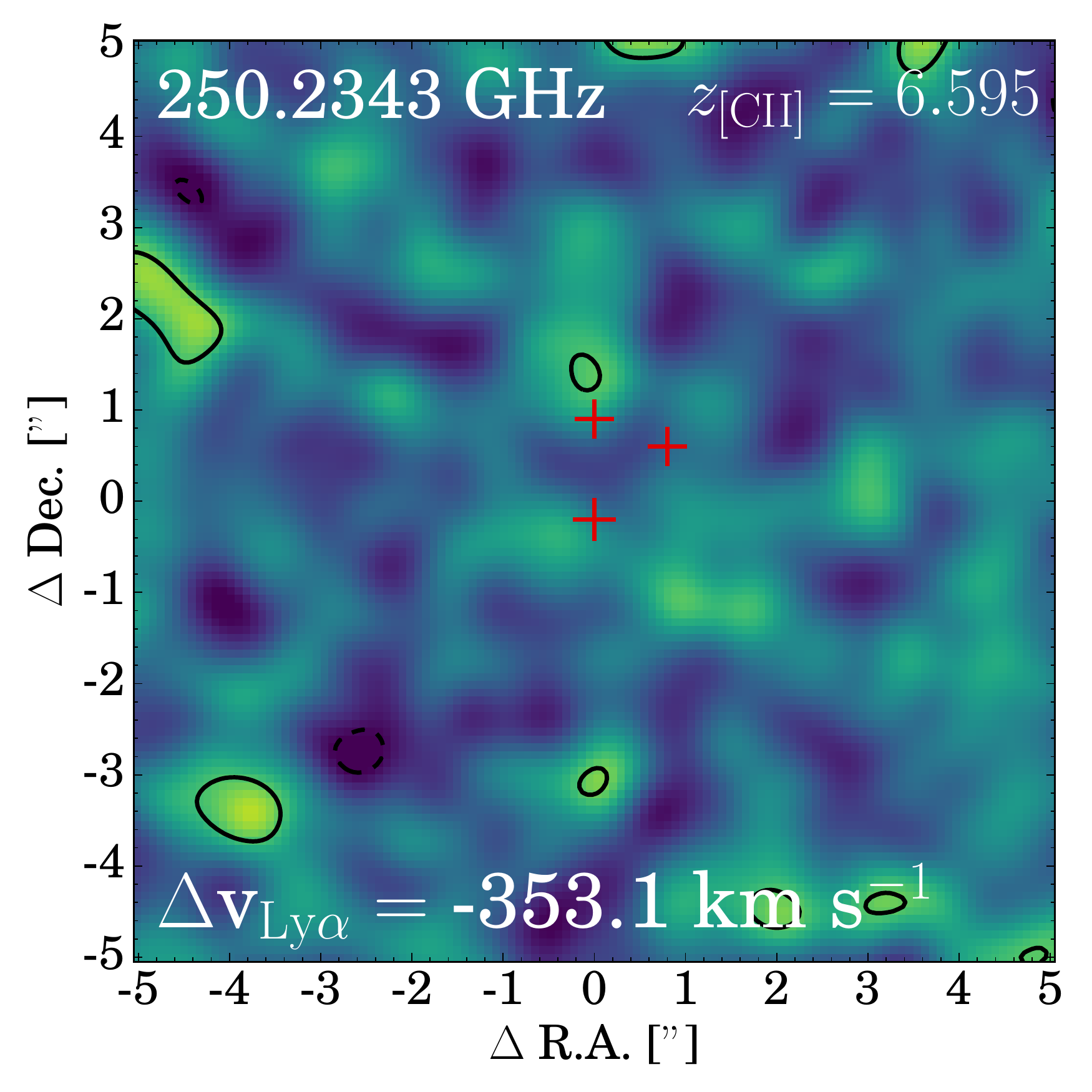}&
	\includegraphics[width=4.2cm]{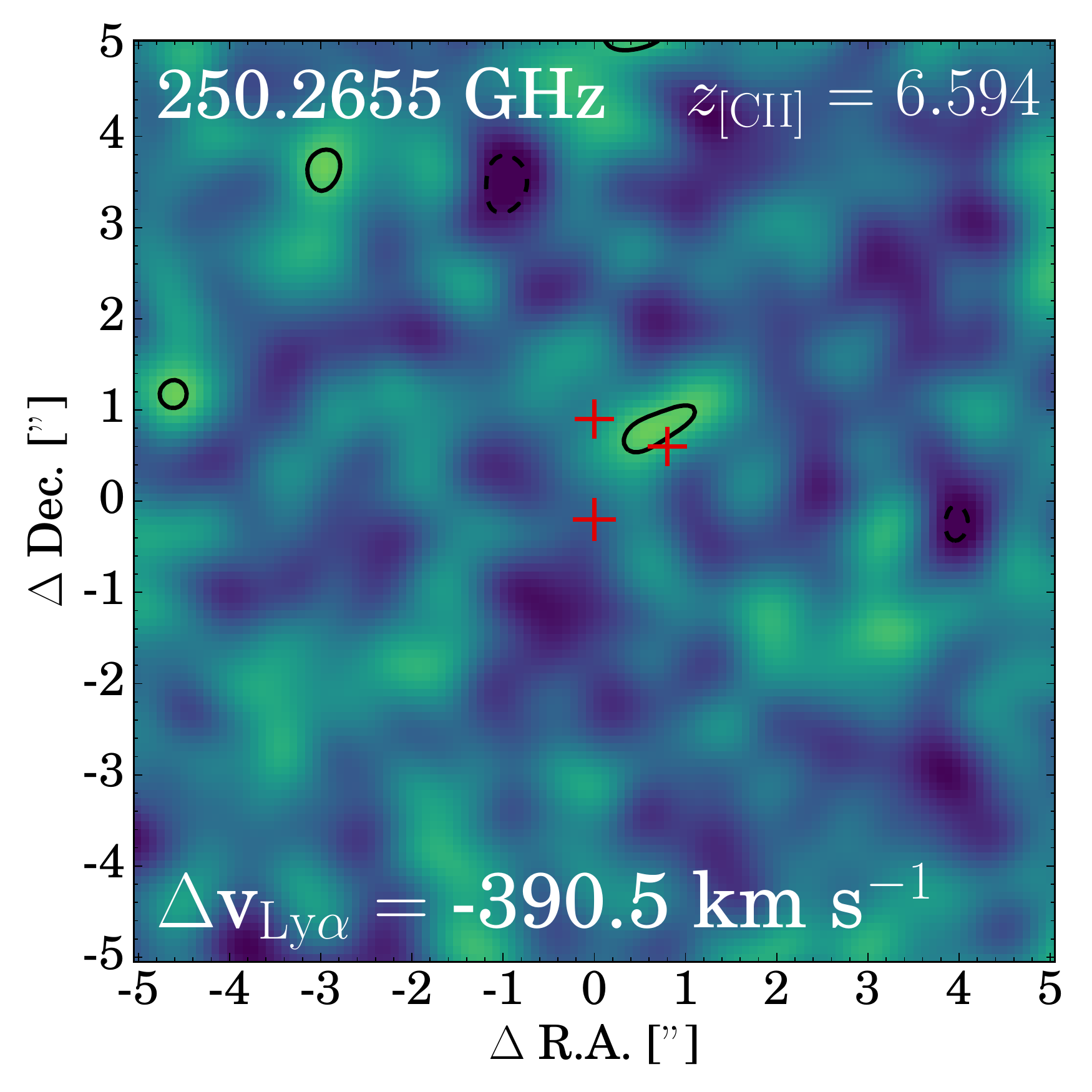}&
	\includegraphics[width=4.2cm]{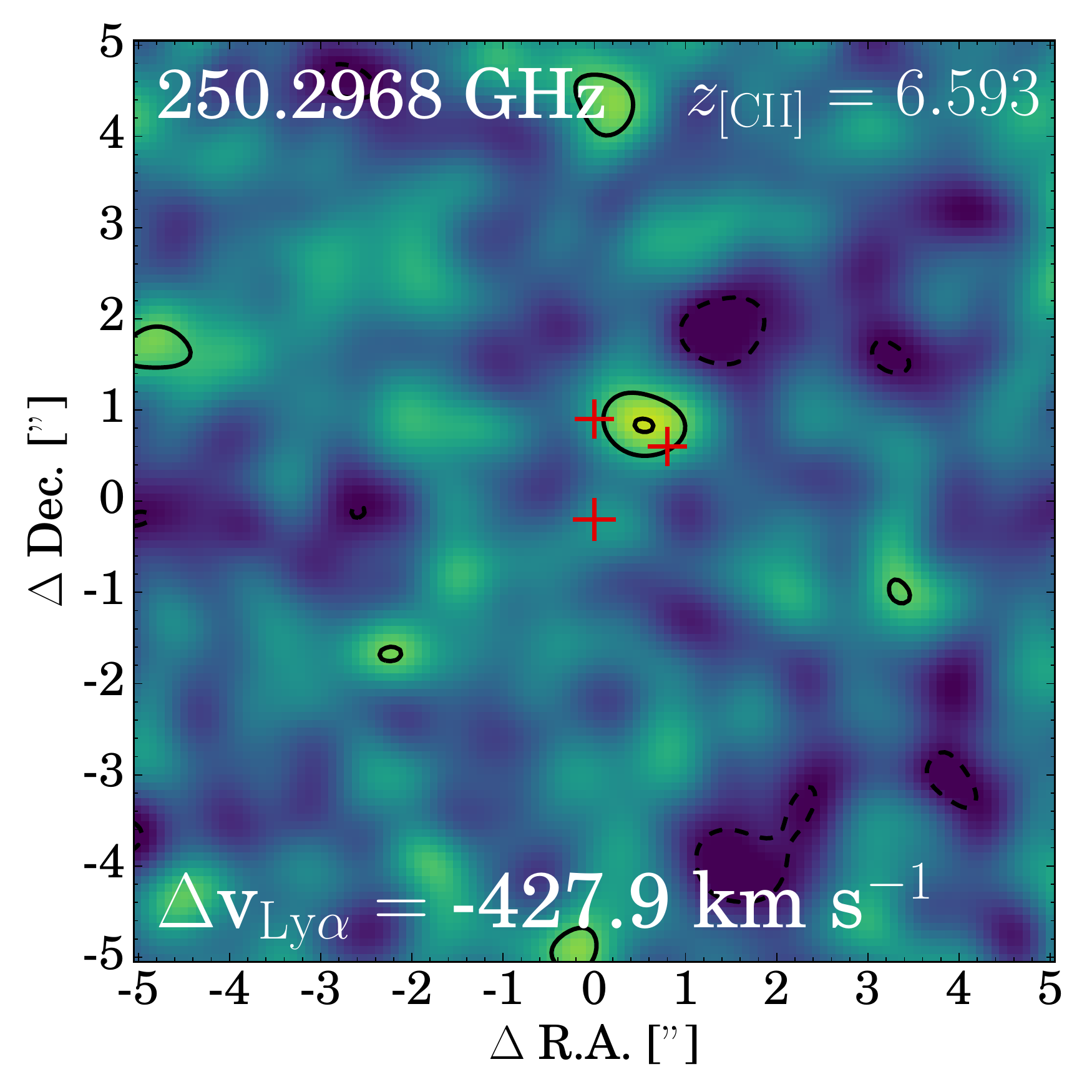}\\
	\includegraphics[width=4.2cm]{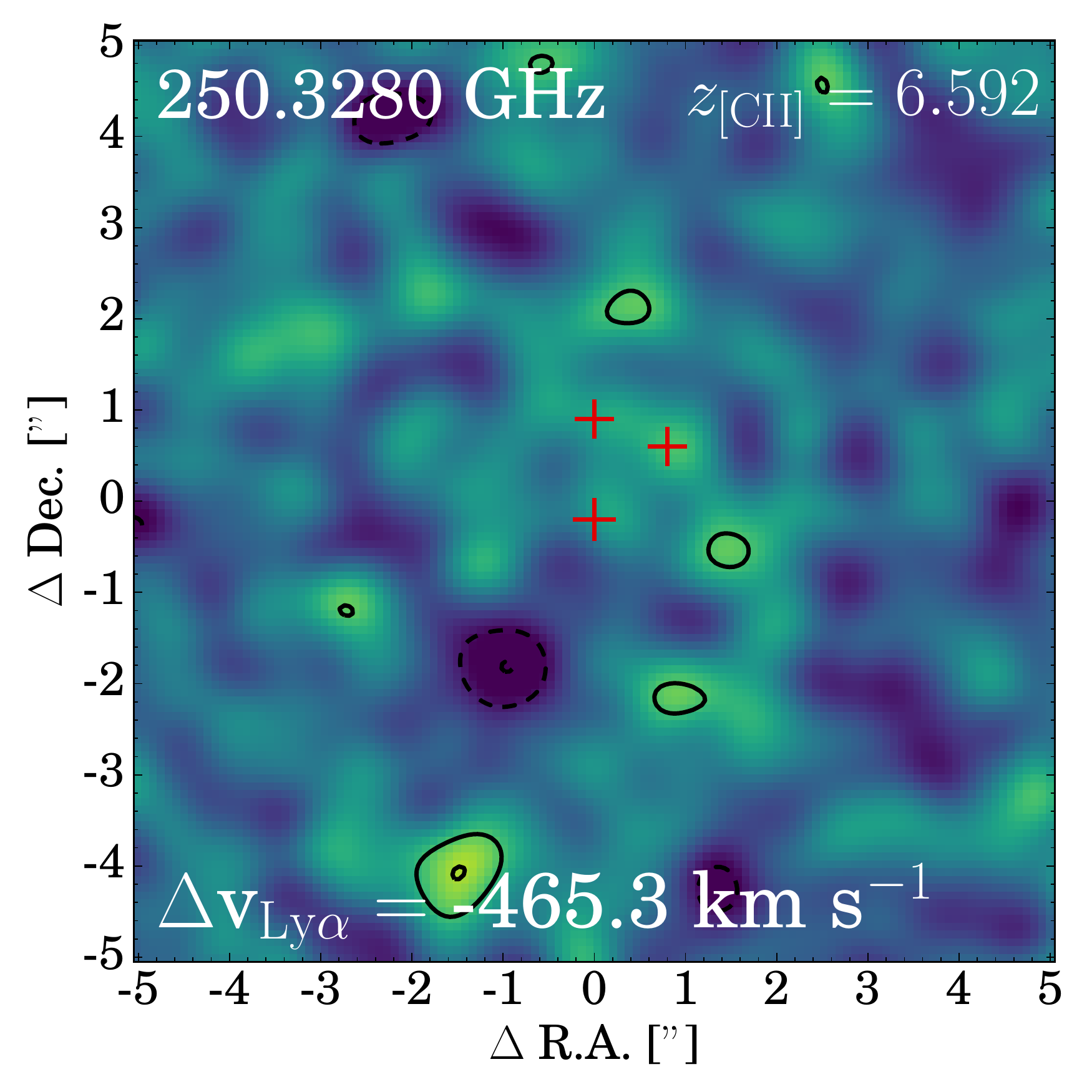}&
	\includegraphics[width=4.2cm]{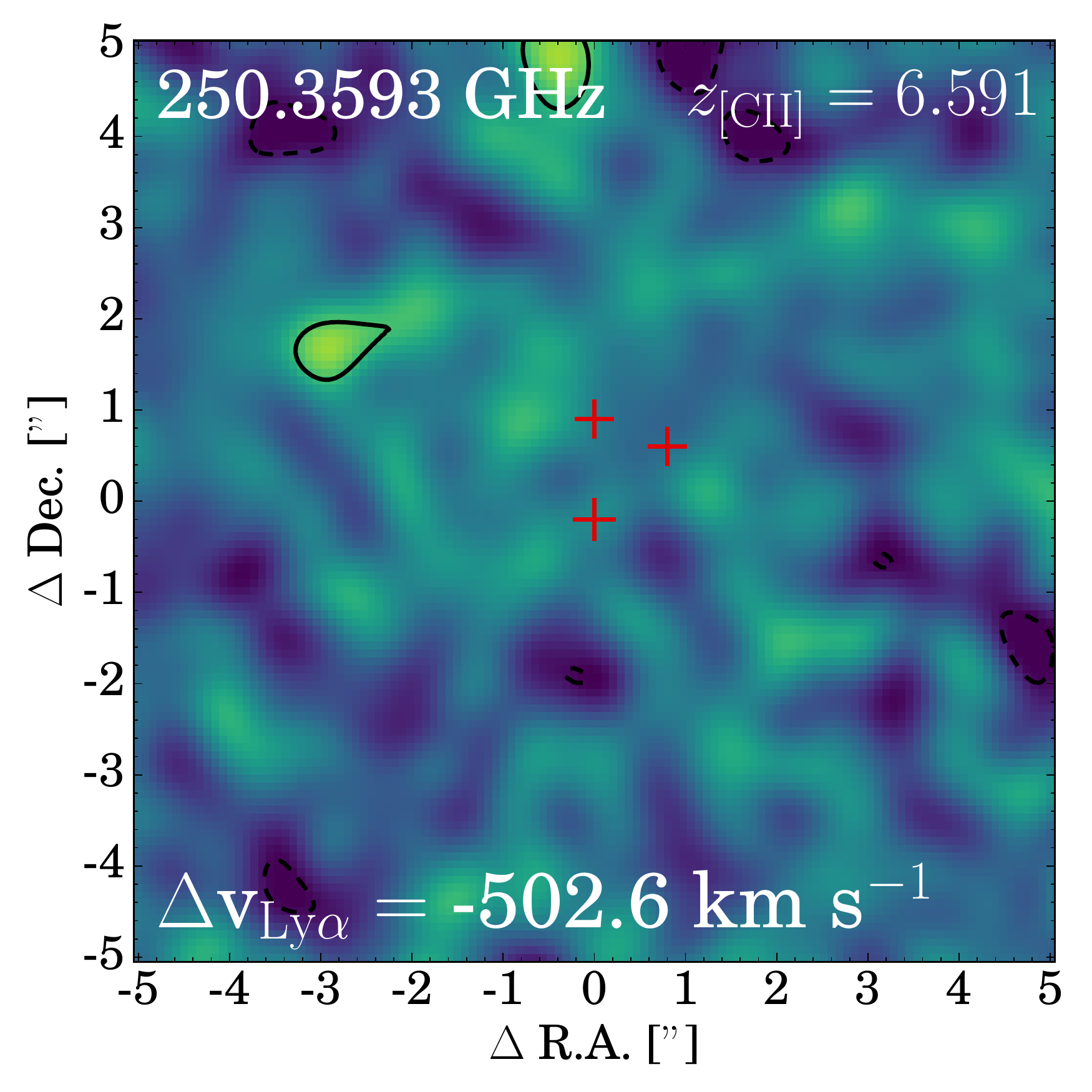}&
	\includegraphics[width=4.2cm]{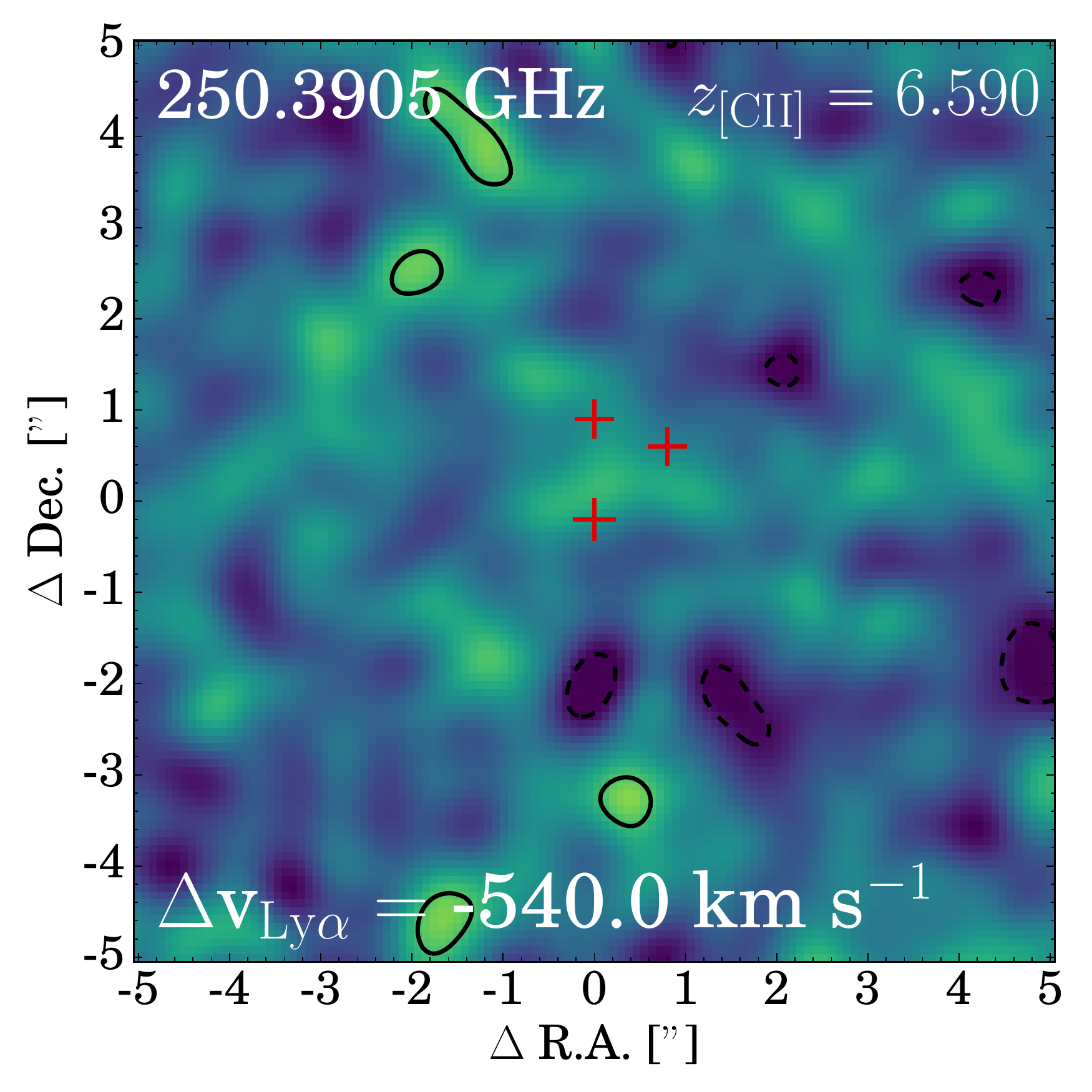}&
	\includegraphics[width=4.2cm]{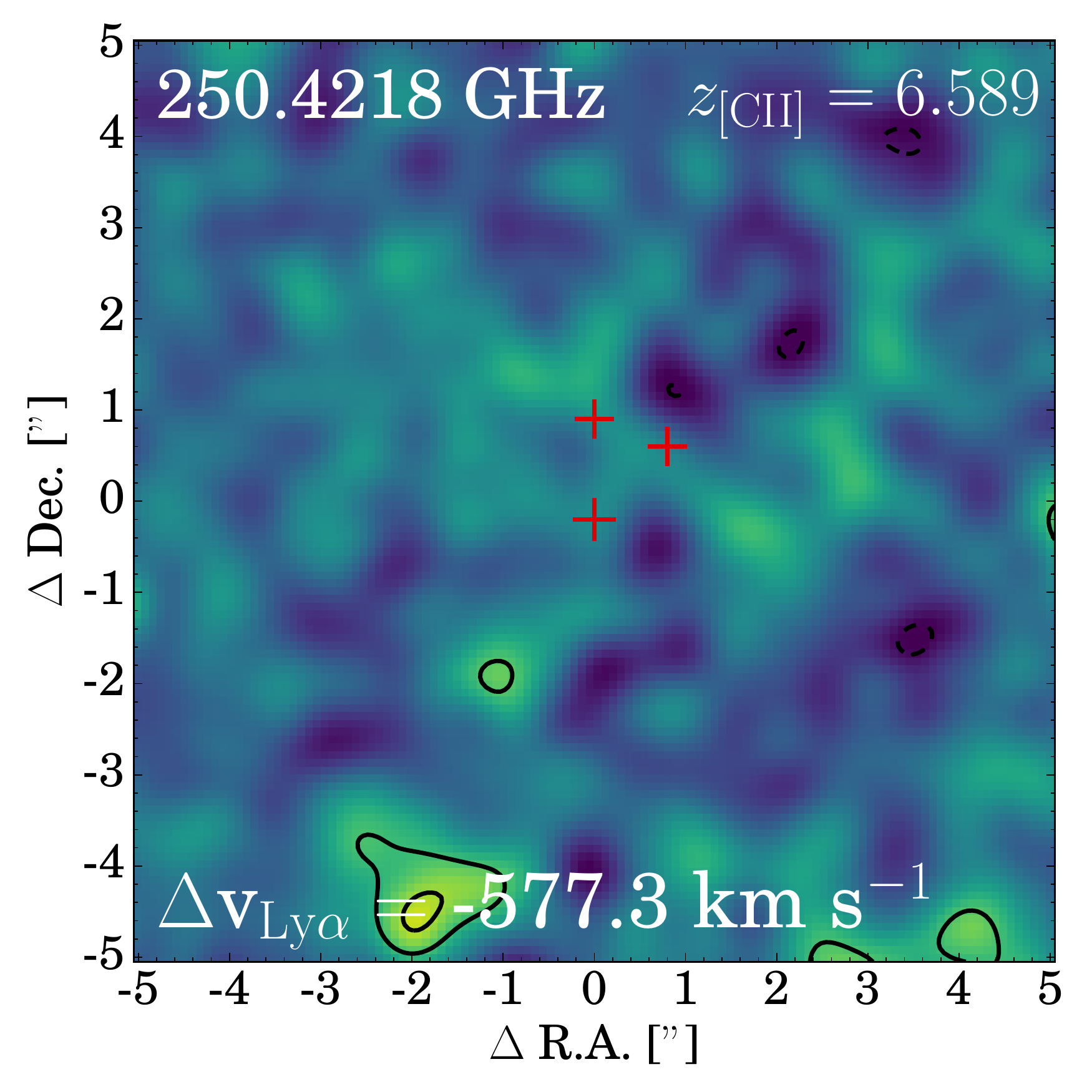}\\
	\end{tabular}
    \caption{Channel maps of the [C{\sc ii}] emission around CR7 ($54\times54$ kpc). Contours show the $\pm 2, 3, 4, 5 \sigma$ levels, where $1\sigma \approx 0.06$ mJy beam$^{-1}$. Red crosses mark the positions of the UV clumps. Channels have widths $\Delta$v$=38.8$ km s$^{-1}$ or 30 MHz. The displayed channels range from $z_{\rm [CII]} = 6.589-6.607$. }
    \label{fig:channels}
\end{figure*}

In order to search for [C{\sc ii}] emission around CR7, we inspect the data-cube of the spectral tuning centered at the frequency where [C{\sc ii}] is expected to be observed. As shown in Fig. $\ref{fig:ALMA}$, resolved [C{\sc ii}] emission is clearly detected.\footnote{The individual [C{\sc ii}] channel maps with widths $\Delta$v$=38.8$ km s$^{-1}$ and frequencies from 249.83-250.42 GHz (corresponding to $z_{\rm [CII]} = 6.607-6.589$) are shown in Fig. $\ref{fig:channels}$.} We show the collapsed data-cube with velocities between $-614$ and $+96$ km s$^{-1}$ with respect to the Ly$\alpha$ redshift, which includes the full velocity range over which we detect [C{\sc ii}] emission. Within the 3$\sigma$ significance contours, we measure L$_{\rm [CII]}$ = $2.17\pm0.36\times10^8$ L$_{\odot}$ in the collapsed image.\footnote{This error has been estimated by measuring the flux in randomly located apertures with the same geometry as the 3$\sigma$ contours.} The spectrum extracted over this region is shown in Fig. $\ref{fig:CII_spectrum}$ and is best-fitted with a line-width of $v_{\rm FWHM}=299\pm27$ km s$^{-1}$ that is offset by $-167\pm27$ km s$^{-1}$ with respect to the Ly$\alpha$ redshift, see Table $\ref{tab:measurements}$. The 3$\sigma$ significance contours span a total area of 2.2 arcsec$^2$ and have a maximum diameter of 1.8$''$, which corresponds to 9.7 kpc at $z=6.60$. This detection of [C{\sc ii}] emission in a large region in/around CR7 clearly indicates the presence of metals, ruling out the possibility that a large fraction of the gas is primordial.

\begin{figure}
	\includegraphics[width=8.6cm]{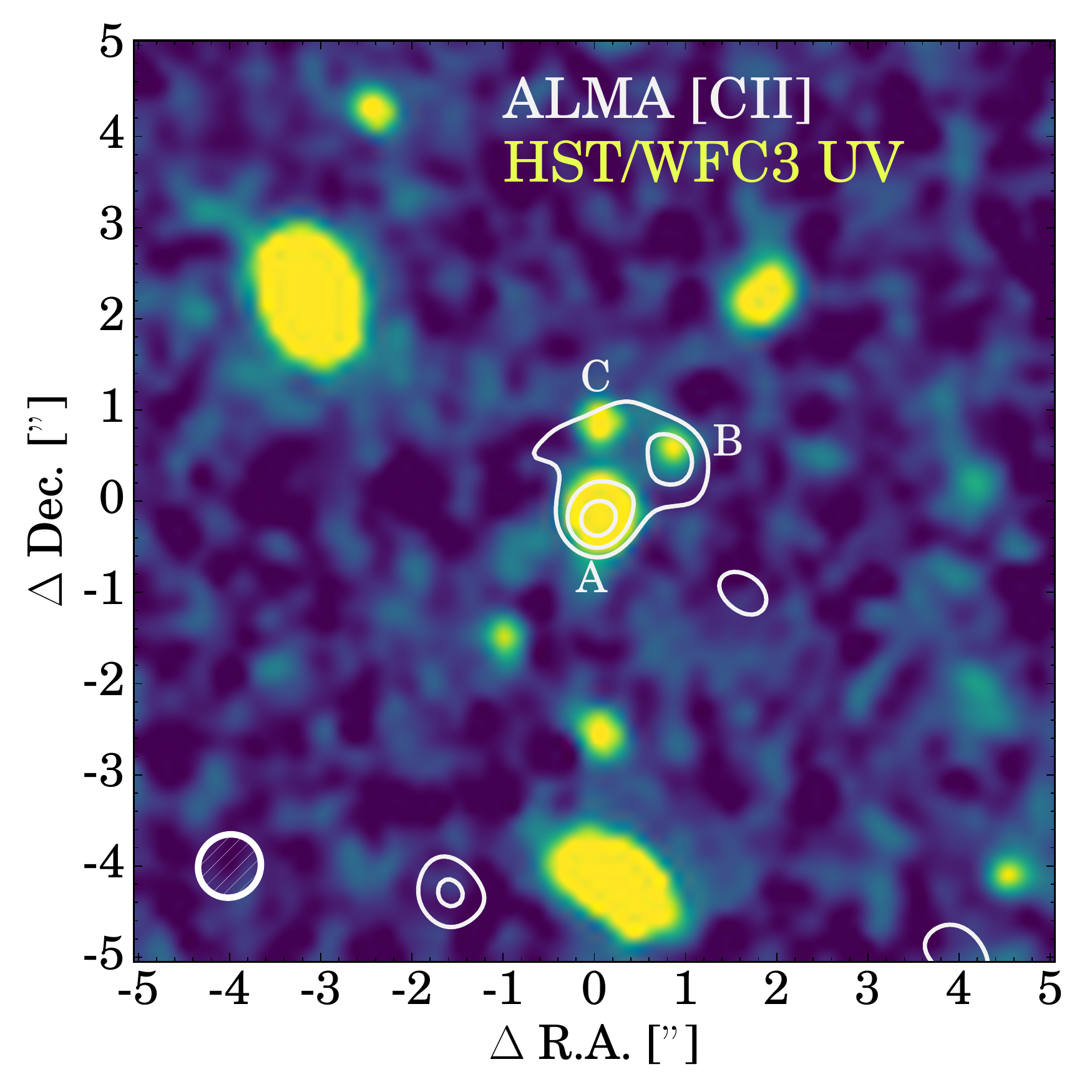}
\caption{Rest-frame UV image (from {\it HST}/WFC3 F110W imaging) of CR7, overlaid with ALMA [C{\sc ii}] 3, 4, 5$\sigma$ contours at a scale of $54\times54$ kpc. The names of the three UV clumps are annotated in HST imaging. Relatively compact components of [C{\sc ii}] emission are detected around clumps A and B, while there is diffuse [C{\sc ii}] emission covering clump C, although at lower significance. \label{fig:ALMA_UV}}
\end{figure}

\subsection{Multiple separated [CII] components}\label{velocity_channels}
As illustrated in Fig. $\ref{fig:ALMA}$, the [C{\sc ii}] emission is clearly not homogeneous over the 3$\sigma$ contour and reveals already two local peaks, resembling the rest-frame UV morphology as observed by {\it HST} (see Fig. $\ref{fig:ALMA_UV}$). In order to show structure in both the spatial and spectral dimensions, we create collapsed [C{\sc ii}] maps in three different velocity ranges of 100 km s$^{-1}$ each and show these in Fig. $\ref{fig:velocity_2ds}$. As the noise level varies slightly at different velocities and is smaller for images that are collapsed over a smaller velocity range, we compute the noise level in each image independently as described in \S $\ref{error_estimate}$. 

The three velocity channels reveal that the [C{\sc ii}] emission in CR7 consists of multiple, well separated components. Around the systemic velocity ($-130$ km s$^{-1}$ with respect to Ly$\alpha$), two sources are clearly detected at the UV positions of clumps A and B. At a blue-shift of $\approx 250$ km s$^{-1}$ a relatively faint, resolved component is detected in between the three UV clumps, but closest to clump C. No significant UV emission is detected at that ALMA position. We name this component C-2. Another compact sub-component is detected at an even larger blue-shift of $\approx 420$ km s$^{-1}$ close to the position of clump B (well within the beam radius), and hence named B-2. In these narrower velocity slices, all detections are at the $\gtrsim4\sigma$ significance level, see Fig. $\ref{fig:velocity_2ds}$.

Hence, we detect several components of [C{\sc ii}] emission in the CR7 system that are significantly offset in both the spatial and spectral dimension. Before discussing the implications of these detections in \S $\ref{discuss_velo}$, we first measure their luminosities and [C{\sc ii}] line-profiles.

\subsubsection{Measurements of individual clumps} \label{CII_HST}
Except for clumps A and C-2, all clumps are unresolved (see \S $\ref{sec:sizes}$). For unresolved sources, we measure the [C{\sc ii}] luminosity within a circular aperture-region with radius of 0.35$''$ (corresponding to the beam semi-major axis). For clumps A and C-2, we use 0.55$''$ and 0.50$''$ apertures respectively to account for the fact that these sources are slightly extended. For these clumps, we find that these apertures retrieve the same fraction of the flux as is retrieved for modelled point sources with a 0.35$''$ aperture. Measurements at {\it HST}/UV positions are done on the collapsed image over the full velocity range (e.g. Fig. $\ref{fig:ALMA}$), while measurements at ALMA positions (B-2 and C-2) are done in the collapsed velocity range in which these clumps are detected (e.g. Fig. $\ref{fig:velocity_2ds}$). 

\begin{figure}
\includegraphics[width=8.6cm]{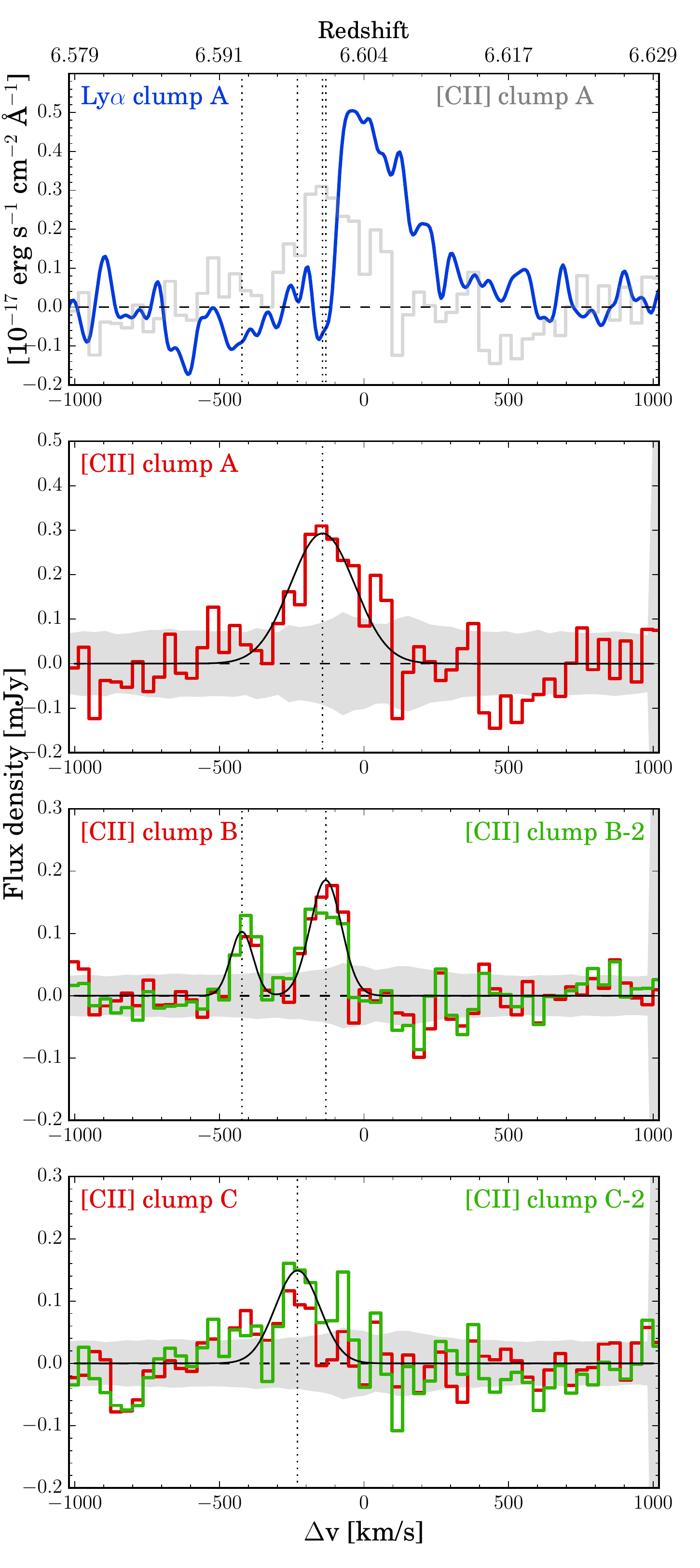}
\caption{Flux densities of Ly$\alpha$ (from X-SHOOTER, centered at clump A) and [C{\sc ii}] at the HST positions of clumps A, B and C as a function of velocity with respect to the Ly$\alpha$ peak velocity at $z=6.604$. We also measure [C{\sc ii}] line-profiles at the positions B-2 and C-2 (shown in green), based on [C{\sc ii}] detections, see Fig. $\ref{fig:velocity_2ds}$. Gaussian fits are shown in black, and we highlight the central velocities with vertical dotted lines. Errors have been estimated for each velocity-slice and aperture as described in \S $\ref{error_estimate}$.  \label{fig:velocities}}
\end{figure}

In Fig. $\ref{fig:velocities}$ we show the ALMA spectra of the different clumps (extracted over their respective apertures) in the rest-frame of clump A ($z=6.601$) and compare these to the Ly$\alpha$ profile measured with VLT/X-SHOOTER at the peak of Ly$\alpha$ emission (close to clump A). We measure the velocity offset between [C{\sc ii}] and Ly$\alpha$ and the [C{\sc ii}] line-width and amplitude by fitting gaussian profiles. All measurements are summarised in Table $\ref{tab:measurements}$.

\begin{figure}
	\includegraphics[width=8.6cm]{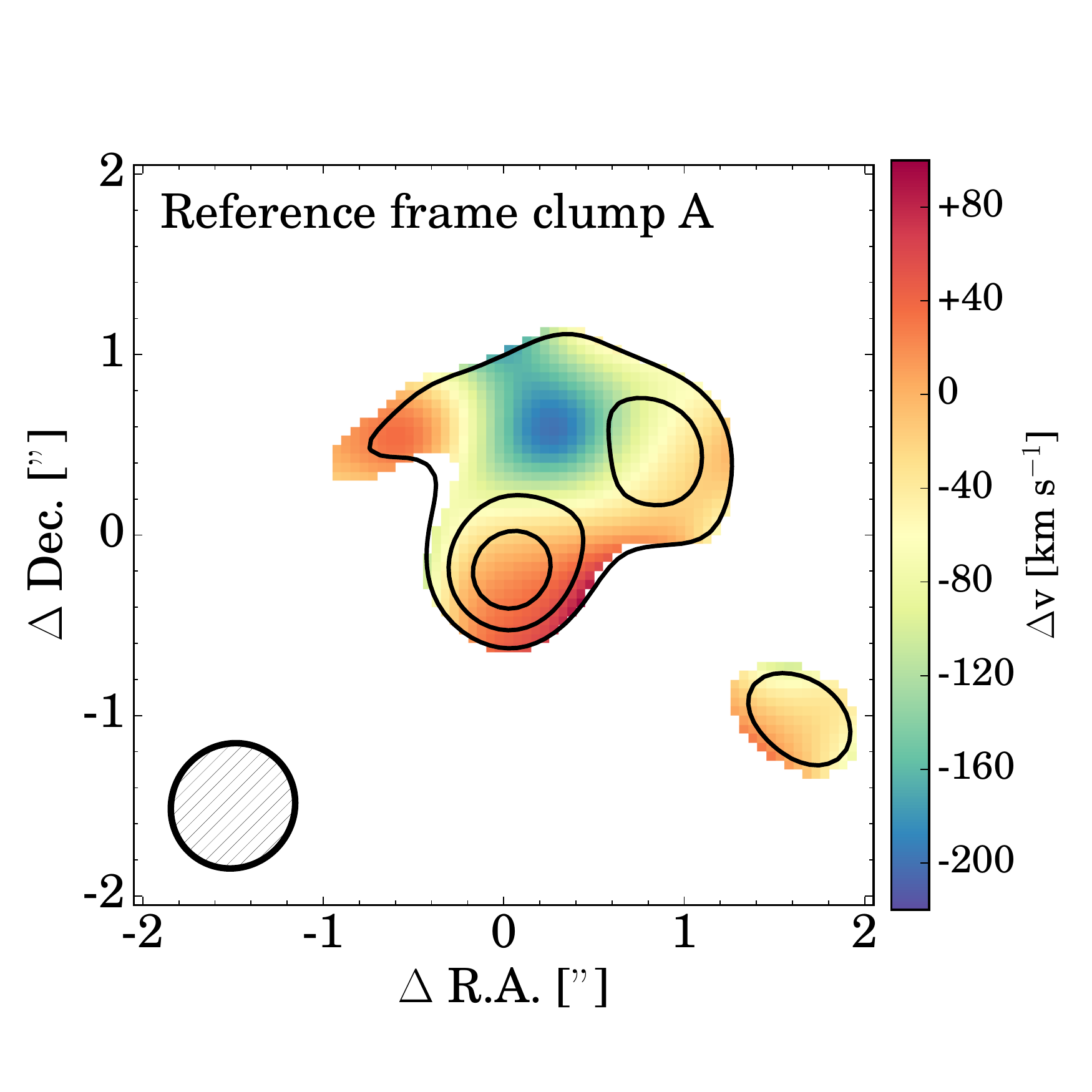}
\caption{[C{\sc ii}] velocity map (at a $22\times22$ kpc scale) in the rest-frame of clump A, based on the first moment map collapsed for frequencies with $-614$ to $+96$ km s$^{-1}$ with respect to the Ly$\alpha$ redshift (see Fig. $\ref{fig:ALMA}$ for the corresponding [C{\sc ii}] flux map). Contours show the 3, 4 and 5$\sigma$ threshold and velocity maps are shown for $>3\sigma$ detections. The map is driven by the the strongly blue-shifted component C-2.    \label{fig:velocitymap}}
\end{figure}

We find that clump A, which is the brightest in the UV and closest to the peak of Ly$\alpha$ emission, is also the brightest in [C{\sc ii}] line-emission, with L$_{\rm [CII]} = (0.96\pm0.20) \times 10^8$ L$_{\odot}$. We caution that we measure a factor two lower luminosity if we measure the luminosity of clump A with a beam-sized aperture. Clump A is blue-shifted by $146\pm27$ km s$^{-1}$ with respect to Ly$\alpha$. This is slightly smaller than typical offsets between Ly$\alpha$ and nebular lines for sources with similar UV luminosities \citep[e.g.][]{Stark2017}, although \cite{Carniani2017} shows a wide spread of velocity offsets in a compilation of ALMA detections (i.e. $-150<\Delta \rm v<+400$ km s$^{-1}$ for galaxies with $10<\rm SFR<100$ M$_{\odot}$ yr$^{-1}$). The [C{\sc ii}] line-width is $259\pm24$ km s$^{-1}$, similar to the Ly$\alpha$ line-width (see e.g. Fig. $\ref{fig:velocities}$). 

At the UV position of clump B, we measure [C{\sc ii}], with L$_{\rm [CII]} = (0.34\pm0.09) \times 10^8$ L$_{\odot}$, which contains the contribution from both the ALMA detections B and B-2 (with $\approx33$ \% of the flux being due to the sub-component B-2, see Table $\ref{tab:measurements}$). The brightest [C{\sc ii}] component of clump B is at $z=6.600\pm0.001$, similar to the redshift of clump A, but has a narrower line-width. Clump B-2 is blue-shifted by $290\pm30$ km s$^{-1}$ with respect to the systemic redshift and has a relatively narrow-line width of $92\pm22$ km s$^{-1}$.

\begin{table*}
\centering
\caption{Properties of different components in CR7. Clumps B-2 and C-2 are based on [C{\sc ii}] detections, while A, B and C are based on {\it HST} detections (see \S $\ref{CII_HST}$). The `Full' uses ground-based imaging and the 3$\sigma$ contour levels from Fig. $\ref{fig:ALMA}$. $\Delta$R.A.$_{\rm HST}$ and  $\Delta$Dec.$_{\rm HST}$ are the difference between the {\it HST} and ALMA positions before shifting the ALMA R.A. by 0.25$''$ to the east. For B-2 and C-2 the differences are between {\it HST} clumps B and C, and ALMA detections B-2 and C-2, respectively. M$_{1500}$ is estimated from F110W imaging after correcting for the contribution of Ly$\alpha$ emission. SFRs assume a Salpeter IMF following \citet{Kennicutt1998}. Dust-attenuation of the UV SFR is estimated based on the UV slope $\beta$ following \citet{Meurer1999}. We provide [C{\sc ii}] luminosities measured from the collapsed image in a beam-sized aperture and [C{\sc ii}] luminosities obtained from the fitted gaussian line-profiles. Dynamical masses are estimated following \citet{Wang2013}.  }
\begin{tabular}{lrrrrrr}
 & Clump A & Clump B & Clump B-2 & Clump C & Clump C-2 & Full \\ \hline
 $\Delta$R.A.$_{\rm HST}$ & $-0.25''$  & $-0.24''$ & $-0.07''$  & - &$+0.14''$ & - \\
 $\Delta$Dec.$_{\rm HST}$ & $-0.12''$ & $-0.18''$ & $-0.04''$ & - &$+0.40''$ & - \\
 M$_{1500}$ & $-21.6\pm0.1$ & $-19.6\pm0.2$ & - & $-20.1\pm0.1$ & $>-19.5$ & $-22.2\pm0.1$ \\ 
 $\beta$ & $-2.3\pm0.4$ &  $-1.0\pm1.0$ & - & $-2.3\pm0.8$ & - & $-2.2\pm0.4$ \\
 SFR$_{\rm UV, no dust}$/M$_{\odot}$ yr$^{-1}$ & $27\pm1$  & $4\pm1$ & - & $7\pm1$ & $<4$  & $44\pm2$\\
 SFR$_{\rm UV, Meurer}$/M$_{\odot}$ yr$^{-1}$ & $29^{+23}_{-2}$ & $38^{+182}_{-32}$ & - & $7^{+19}_{-1}$ & - & $50^{+55}_{-6}$ \\
 SFR$_{\rm UV+IR}$/M$_{\odot}$ yr$^{-1}$ & $28^{+1}_{-1}$ & $5^{+2}_{-1}$ &  - & $7^{+1}_{-1}$ & $<4$ & $45^{+2}_{-2}$ \\

 $\Delta$v$_{\rm Ly\alpha}$/km s$^{-1}$ & $-146\pm27$ & $-152\pm21$ & $-422\pm22$ & - & $-232\pm27$ & $-167\pm27$ \\ 
 $z_{\rm [CII]}$ & $6.601\pm0.001$ & $6.600\pm0.001$ & $6.593\pm0.001$ & - & $6.598\pm0.001$ & $6.600\pm0.001$ \\
 $v_{\rm FWHM, [CII]}$/km s$^{-1}$ & $259\pm24$ & $130\pm21$ &$92\pm22$ & - & $181\pm30$ & $299\pm26$ \\  
 
 S$_{\nu, \rm fit}\Delta$v /mJy km s$^{-1}$ & $81.1\pm20.7$ & $25.0\pm5.3$ & $10.0\pm3.9$ & - & $35.4\pm10.9$ & $185.2\pm39.0$  \\
  L$_{\rm [C{\sc II}], fit}$/10$^8$ L$_{\odot}$ & $0.88\pm0.23$ & $0.26\pm0.07$  &$0.12\pm0.03$ & -   & $0.27\pm0.11$ & $2.00\pm0.43$  \\

 L$_{\rm [C{\sc II}], aperture}$/10$^8$ L$_{\odot}$ & $0.96\pm0.20$ & $0.34\pm0.09$  &$0.11\pm0.03$   & $0.26\pm0.09$ & $0.27\pm0.07$ & $2.17\pm0.36$  \\
 
 r$_{1/2, \rm [C{\sc II}]}$/kpc & $3.0\pm0.8$ & $<2.2$ & $<2.2$ & - & $3.8^{+1.1}_{-0.7}$ & - \\ 
  M$_{\rm dyn}$/$(\sin i)^2$ 10$^{10}$ M$_{\odot}$ & $3.9\pm1.7$ & $<0.7$ & $<0.4$ & - & $2.4\pm1.9$ & - \\ \hline   
   
 L$_{\rm UV}$/10$^{10}$ L$_{\odot}$ & 9.1 & 1.4 & - & 2.3 & $<1.3$ & 15.8 \\
rms$_{\rm cont, \lambda_0 = 160 \mu m}$/$\mu$Jy beam$^{-1}$ & 7 & 7 & 7 & 7 & 7 & 7 \\  
 L$_{\rm IR}$ ($T_{d} = 35$K)/10$^{10}$L$_{\odot}$ & $<3.14$ & $<3.14$ & $<3.14$ & $<3.14$ & $<3.14$ & $<3.14$  \\  
 SFR$_{\rm IR}$ ($T_{d} = 35$K)/M$_{\odot}$ yr$^{-1}$ & $<5.4$ & $<5.4$ & $<5.4$ & $<5.4$ & $<5.4$ & $<5.4$  \\  
 
 M$_{\rm dust}$($T_{d} = 35$K)/10$^6$M$_{\odot}$ & $<8.1$ & $<8.1$ & $<8.1$ & $<8.1$ & $<8.1$ & $<8.1$ \\
 
\hline
\label{tab:measurements}
\end{tabular}
\end{table*}

At the position of clump C, we measure relatively faint [C{\sc ii}] emission at the $\approx3\sigma$ level, see Fig. $\ref{fig:ALMA}$ and Fig. $\ref{fig:ALMA_UV}$. The aperture at the position of clump C is contaminated slightly by this clump B-2 (which may also be associated with C), explaining the bluest line, but the redder [C{\sc ii}] component C-2 is offset spatially (see Fig. $\ref{fig:velocity_2ds}$). At the position of C-2, we measure L$_{\rm [CII]}$=$(0.27\pm0.07) \times10^8$ L$_{\odot}$ with a line-width of $181\pm30$ km s$^{-1}$ and an offset of $-232\pm27$ km s$^{-1}$ with respect to Ly$\alpha$. This offset is bluer than clumps A and B, but redder than B-2. Due to its large blue-shift, component C-2 also dominates the velocity map shown in Fig. $\ref{fig:velocitymap}$. Due to this strong contribution of component C-2 to the velocity map, we find that there is no clear evidence for ordered rotation over the full [C{\sc ii}] extent. This shows that the spatial resolution of our observations is crucial, as we may have not been able to distinguish clumps C-2 and B/B-2 with lower resolution, and could have mis-interpreted the velocity map as rotation. It is challenging to investigate whether ordered rotation is present inside individual clumps because they are either unresolved or only marginally resolved. Similarly as for clump A, we note that the [C{\sc ii}] luminosity of clump C-2 would be reduced by a factor $\approx2$ if we use a smaller aperture of 0.35$''$, which is used for unresolved sources. Therefore, the luminosities of clumps A and C-2 should be interpreted with caution. The total [C{\sc ii}] luminosity that is associated to clumps is $(1.57\pm0.23)\times10^8$ L$_{\odot}$, which is the sum of the aperture measurements at the {\it HST} positions. This corresponds to a fraction of $0.72\pm0.18$ of the total observed [C{\sc ii}] luminosity.

\subsection{Sizes and dynamical masses} \label{sec:sizes}
Starting from the measured velocity width and size of the emitting regions, it is possible to obtain a rough estimate of the dynamical mass of the various [C{\sc ii}] components \citep[e.g.][]{Wang2013}. First, we measure the size by performing two dimensional gaussian fits in the collapsed 2D image. Clumps A and B are simultaneously fitted in the image that is collapsed over the full velocity range, while clumps B-2 and C-2 are fitted in the collapsed image in the velocity range as shown in Fig. $\ref{fig:velocity_2ds}$. In our fitting procedure, we fixed the position angle and the ratio between the dispersion in both dimensions corresponding to the those of the beam semi-major and semi-minor axis. We also fix the positions to those indicated in Fig. $\ref{fig:velocity_2ds}$. Hence, the free parameters are the amplitudes and widths of the gaussian profiles. In order to take the (correlated) noise into account, we add the noise from a randomly selected cut-out region in the collapsed image, avoiding the source itself, before performing the fit. This is repeated 5000 times in order to estimate the statistical reliability of the fit. We find that more than $>50 \%$ of the measurements for clumps B and B-2 result in a size that is equal to or smaller than the beam, meaning that these clumps are unresolved and hence have a size r$_{1/2, \rm [CII]}<2.2$ kpc. We measure an observed size of r$_{1/2, \rm [CII], obs} = 3.7^{+0.6}_{-0.6}$ kpc for clump A (4 \% of the measurements result in an unresolved size, meaning that clump A is resolved at $\approx2\sigma$ significance) and r$_{1/2, \rm [CII], obs} = 4.4^{+0.9}_{-0.6}$ kpc for clump C-2 (resolved at $\approx3.5\sigma$ significance). We deconvolve the sizes of A and C-2 to obtain the intrinsic size with  $r_{\sigma,\rm [CII]}= \sqrt{r_{\rm obs}^2 - 2.2^2}$. Resulting sizes are listed in Table $\ref{tab:measurements}$. Dynamical masses are computed following \cite{Wang2013}:
\begin{equation}
  \rm M_{\rm dyn} /M_{\odot} (\sin {\it i})^2 = 1.94\times10^5 \times v_{\rm FWHM, [CII]}^2 \times  r_{1/2,\rm [CII]},
\end{equation}
where M$_{\rm dyn}$ is the dynamical mass in M$_{\odot}$, $i$ is the inclination angle, $v_{\rm FWHM, [CII]}$ the line-width in km s$^{-1}$ and $r_{1/2,\rm [C{\sc II}]}$ the size in kpc (half-light radius). This results in dynamical masses (uncorrected for inclination) ranging from $(3.9\pm1.7)\times10^{10}$ M$_{\odot}$ for component A, $(2.4\pm1.9)\times10^{10}$ M$_{\odot}$ for clump C-2 and $<0.7\times10^{10}$ M$_{\odot}$ for $<0.4\times10^{10}$ M$_{\odot}$ for clumps B and B-2, respectively. These dynamical mass estimates are lower than typical quasar host galaxies at $z\approx6$ \citep{Wang2013}, and comparable to star-forming galaxies at $z\approx7$ with similar SFRs as CR7 \citep[e.g.][]{Pentericci2016,Smit2017}.

\begin{figure}
\centering
	\includegraphics[width=8.6cm]{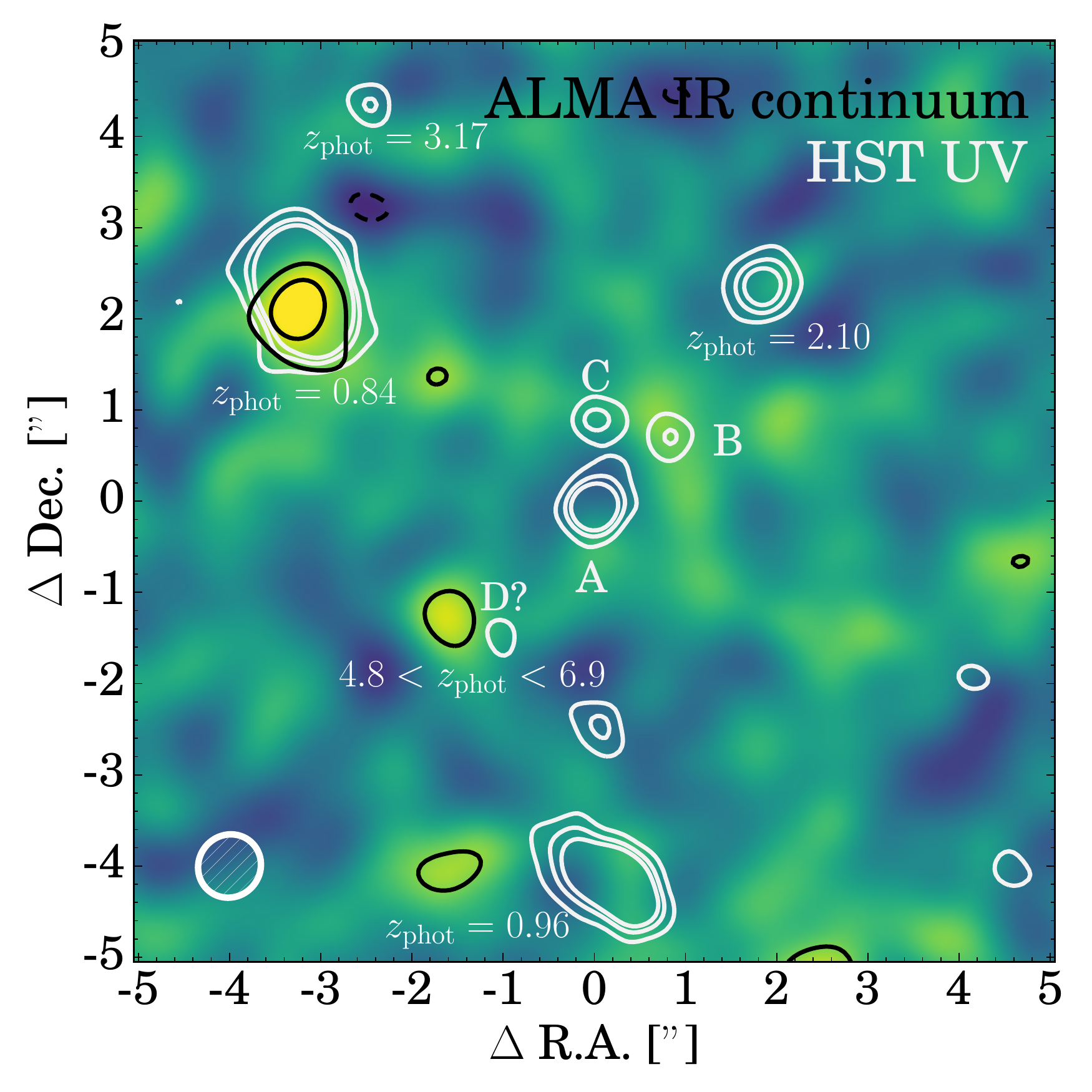}
    \caption{IR continuum map at 230-250 GHz centered on the position of CR7. The black contours show the 2, 3$\sigma$ level, where 1$\sigma = 7 \mu$Jy beam$^{-1}$. We also show {\it HST} rest-frame UV (F814W+F110W+F160W) contours at the 2, 3, 4$\sigma$ levels to highlight the positions of known (foreground) sources. Sources are annotated with their photometric redshifts estimated by \citet{Laigle2016}. No dust continuum is detected at CR7, although a $\approx2\sigma$ signal is detected around potential clump D. Dust continuum is also clearly detected in a foreground source at $z=0.84$. } 
    \label{fig:continuum_map}
\end{figure}

\section{IR continuum} \label{sec:FIR}
\subsection{Blind detections} \label{sec:FIR_blind}
We combine the flux in all four spectral windows from our ALMA coverage to search for dust continuum emission. In the entire image, we find two detections with S/N $>3$, but they are not associated with CR7. One detection is 3.5$''$ north-east of CR7 (associated with ID number 339509 in the catalog from \citealt{Laigle2016}, photo$-z=0.84$ and visible in Fig. $\ref{fig:continuum_map}$), while the other is 18.5$''$ to the south-west (ID number 335753 in \citealt{Laigle2016}, $photo-z=3.10$, not visible in the image). The positions of these foreground galaxies confirm the astrometric correction described in \S $\ref{sec:astrometry}$. We note that we detect a tentative (3$\sigma$) line at 250.484 GHz at the position of ID 339509 that is identified as CO(4-3) with $\nu_0 = 461.041$ GHz at $z=0.841$, perfectly consistent with its photometric redshift.

As visible in Fig. $\ref{fig:continuum_map}$, there is a $\approx2.7\sigma$ continuum detection $\approx2''$ to the south-east of CR7, nearby a faint {\it HST} detection that we name potential clump D. This faint {\it HST} detection has F110W=$27.2\pm0.2$ in a 0.4$''$ aperture, similar to clump B. It is also detected at $\approx3\sigma$ in the ACS/F814W filter (F814W$=27.5\pm0.2$) and F160W filter (F160W$=27.0\pm0.3$). It is not detected in NB921 and $z'$ imaging, which have 3$\sigma$ depths of 25.8 AB magnitude, hence consistent with the {\it HST} photometry. Potential clump D is not detected in the Subaru S-cam imaging in the $B, V, R$ filters with 2$\sigma$ upper limits of $\lesssim 28.5-29.0$. Therefore, the photometry is consistent with the Lyman-break falling in the F814W filter, such that $4.8<z_{\rm phot}<6.9$. There is no significant [C{\sc ii}] detection at its position in any of the spectral tunings observed with ALMA. Further IFU observations are required to measure the redshift of clump D and test whether it is associated to CR7.

\subsection{Upper limits for CR7}
There is no IR continuum detected at the position of CR7's clumps A, B and C, see Fig. $\ref{fig:continuum_map}$. Around the position of CR7, we measure an rms of 7 $\mu$Jy beam$^{-1}$ at $\lambda_0 \approx 160 \mu$m.  Note that we do not remove contamination from the [C{\sc ii}] line as we expect that this is negligible, as a result our upper limit is therefore somewhat on the conservative side. We follow \cite{Schaerer2015} to convert this limit into a 3$\sigma$ upper limit on the infrared luminosity (between 8-1000 $\mu$m) of L$_{\rm IR} < 3.14\times10^{10}$ L$_{\odot}$, under the assumption that the dust temperature is 35 K and a modified black body SED with a power-law slope of $2.9$ in the Wien regime, $\beta_{\rm IR} = 1.5$ and after removing the contribution of the CMB to the dust heating, see \cite{daCunha2013}. In the case that the dust temperature is 25 (45) K, the limit on the IR luminosity is L$_{\rm IR} < 1.68 (6.15) \times10^{10}$ L$_{\odot}$. Combined with the total [C{\sc ii}] luminosity, we find a lower limit of the the ratio log$_{10}$(L$_{\rm [CII]}$/L$_{\rm IR}$)$>-2.1$, which is similar to the highest values measured in local star-forming galaxies \cite[e.g.][]{Malhotra2001} and significantly higher than sub-mm galaxies and quasar host galaxies at $z>3$ (see \citealt{Ota2014} for a compilation). This could potentially indicate a lower dust-to-metal ratio, which in turn could result in a higher dust temperature due to the lower dust opacity \citep[e.g.][]{CenKim2014}.

\begin{figure*}
\centering
	\includegraphics[width=15cm]{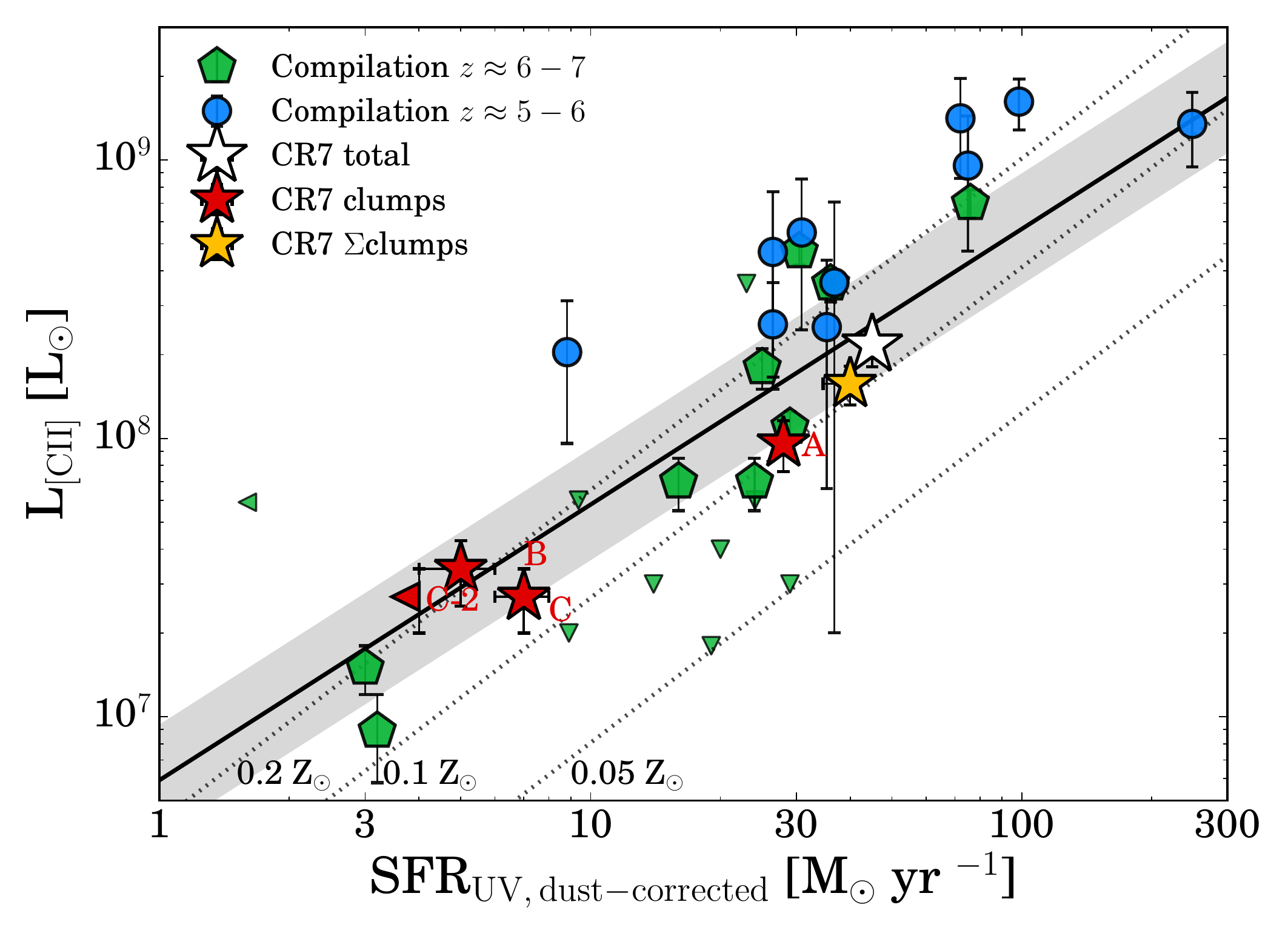}
    \caption{Dust corrected SFR(UV) versus [C{\sc ii}] luminosity for a compilation of galaxies at $z\sim6$ (blue circles; \citealt{Capak2015,Willott2015}) and $z\sim7$ (green pentagons; \citealt{Kanekar2013,Ouchi2013,Ota2014,Maiolino2015,Schaerer2015,Knudsen2016,Pentericci2016,Bradac2017,Knudsen2017,Smit2017}). We show the individual UV components of CR7 with red stars, and the sum of these component as an orange star. The integrated measurement over the full [C{\sc ii}] emitting region around CR7 is shown with an open star. Downward pointing triangles show galaxies with upper limits on the [C{\sc ii}] luminosity, while left-ward pointing triangles show [C{\sc ii}] detections with upper limits on SFR(UV). The black line shows the relation for local star-forming galaxies measured by \citet{DeLooze2014}, where the grey region shows the observed dispersion. The dotted lines are model-predictions from \citet{Vallini2015} on how the [C{\sc ii}] luminosity varies with metallicity. We find that the clumps in CR7 are within the scatter from the local relation. These measurements indicate metallicities $0.1<$ Z/Z$_{\odot}<0.2$. Note that we have corrected published SFRs (including those from the local relation) to the Salpeter IMF.} 
    \label{fig:SFR_CII}
\end{figure*}

The ratio of the IR$_{160 \mu\rm m}$ to UV$_{0.15 \mu\rm m}$ flux density is constrained to ($\nu \rm F_{\nu})_{\rm IR} / (\nu \rm F_{\nu})_{\rm UV} < 0.04$ (3$\sigma$). This limit is almost an order of magnitude stricter than other $z\sim7$ sources observed by \cite{Maiolino2015}. In the local Universe, this ratio is related to metallicity (see the compilation and fitted relation in \citealt{Maiolino2015}), likely as a result of dust content increasing with metallicity \citep[e.g.][]{RemyRuyer2014}. If such a relation would be unchanged at $z=7$ \citep[e.g.][]{Popping2017}, the measurements imply a metallicity of $\rm 12+log(O/H) \lesssim 7.5$ (corresponding to $\lesssim0.1$ Z$_{\odot}$), which is only found in dwarf galaxies in the local Universe with masses M$_{\rm star} \lesssim 10^8$ M$_{\odot}$ \citep[e.g.][]{Izotov2015,Guseva2017}.

We use the upper limits on the IR continuum to place an upper limit on the dust mass. We follow the method outlined in \cite{Ota2014}, which assumes a dust mass absorption coefficient $k_{\nu} = 1.875 (\nu/239.84)^{\beta_{\rm IR}}$ m$^2$ kg$^{-1}$ where $\nu$ is in GHz and $\beta_{\rm IR} = 1.5$ and removes the contribution from the CMB. This results in M$_{\rm dust} < 8.1 (27.5)\times10^6$ M$_{\odot}$ if the dust temperature is 35 (25) K. These limits are consistent with expectations for CR7's UV luminosity, based on post-processed hydrodynamical simulations by \cite{Mancini2016}, who find dust masses between $\approx1-10\times10^5$ M$_{\odot}$. 

Following the prescription from \cite{Kennicutt1998}, we measure a 3$\sigma$ upper limit of the dust obscured SFR$_{\rm IR} < 5.4$ M$_{\odot}$ yr$^{-1}$ for a \cite{Salpeter1995} IMF. This allows us to put a stronger constraint on the total SFR of different clumps. Because of these tight limits, the large errors in $\beta$ (which for example allow relatively red UV slopes and hence relatively large dust obscurations) are mitigated. Combining these measurements (by constraining that the difference between dust-corrected SFR and dust-free SFR is $<5.4$ M$_{\odot}$ yr$^{-1}$ in each of the 10,000 realisations used to self-consistently compute the uncertainties as described in \S $\ref{sec:UVprops}$) results in SFR$_{\rm UV+IR} = 28^{+1}_{-1},  5^{+2}_{-1},  7^{+1}_{-1}$ M$_{\odot}$ yr$^{-1}$ for clump A, B and C, respectively. For the total (ground-based) photometry, we find SFR$_{\rm UV+IR} =45^{+2}_{-2}$ M$_{\odot}$ yr$^{-1}$, which is roughly consistent with the sum of the {\it HST} detected clumps and indicates no other significant source of star-formation.

\section{The SFR-L$_{\rm [CII]}$ relation} \label{sec:SFRLCII}  
Due to its luminosity, the [C{\sc ii}] fine-structure cooling line has been proposed as a SFR tracer (see e.g. \citealt{DeLooze2014} and references therein), which is observable with ALMA at high-redshift, while other tracers such as H$\alpha$ are currently unavailable. However, while [C{\sc ii}] is insensitive to dust attenuation, unlike (for example) UV emission, its luminosity is sensitive to metallicity \citep[e.g.][]{Vallini2013,Hemmati2017,Olsen2017}. The [C{\sc ii}] luminosity is also dependent on the ionisation state of the gas and saturates at high temperatures \citep{Kaufman1999,Decataldo2017,Vallini2017}.

Early ALMA observations of galaxy samples at $z\approx5-6$ \citep[e.g.][]{Capak2015,Willott2015} found that these galaxies have similar L$_{\rm [CII]}$ to SFR(UV) ratios as local star-forming galaxies \citep{DeLooze2014}. Galaxies at $z\approx6-7$ were typically selected based on their Ly$\alpha$ redshifts and were initially observed at fainter [C{\sc ii}] luminosities or luminosity limits \citep[e.g.][]{Kanekar2013,Ouchi2013,Ota2014,Maiolino2015,Schaerer2015}. This was later confirmed in observations of lensed galaxies \citep{Knudsen2016,Bradac2017}. However, sources that have been selected independently of strong Ly$\alpha$ emission show a wider range in  [C{\sc ii}] luminosities \citep[e.g.][]{Pentericci2016,Knudsen2017,Smit2017}, although some may still be relatively luminous in Ly$\alpha$ \citep[e.g.][]{Laporte2017b}. 

In Fig. $\ref{fig:SFR_CII}$, we show how our [C{\sc ii}] measurements of the UV components of CR7, their sum, and the integrated CR7 measurements compare to samples at $z\sim5-7$ and to the local relation. As we use the aperture measurements centred at the UV clumps, the [C{\sc ii}] luminosity of clump B includes that of sub-component B-2. We also show the upper limit of the UV SFR of clump C-2, if we do not associate its [C{\sc ii}] luminosity with the UV luminosity of clump C. Most sources that currently have been targeted are UV luminous (with SFR $\approx20-30$ M$_{\odot}$ yr$^{-1}$, similar to clump A), while the two lensed sources have much lower SFRs.\footnote{Note that we have also rescaled all published SFRs, including the local relation from \cite{DeLooze2014}, to a Salpeter IMF.} CR7's clumps B and C lie roughly in the middle of the probed parameter space. All individual clumps in CR7, and the total luminosity are consistent within the observed scatter and their error-bars with the local relation, with only a marginal offset. This is in contrast with previously targeted luminous LAEs such as Himiko \citep{Ouchi2013} and IOK-1 \cite{Ota2014}.

Combining our measurements of clump A and CR7s total luminosity with sources from the literature, we confirm a large spread of an order of magnitude in  [C{\sc ii}] luminosities at fixed UV SFR, indicating that these galaxies exhibit a range in metallicities and/or ionisation states (the dotted lines illustrate how the [C{\sc ii}] strength changes with metallicity in the model from \citealt{Vallini2015}). Using these models, we find that the components in CR7 would have gas-metallicities $0.1<$ Z/Z$_{\odot}<0.2$, with relatively little variation between different components, although components A and C are  consistent with being slightly more metal poor than B.

\subsection{Are UV continuum, [CII] and Ly$\alpha$ related?}
Because of the initial non-detections of [C{\sc ii}] emission in galaxies with strong Ly$\alpha$ emission \citep[e.g.][]{Ouchi2013,Ota2014}, it has been speculated that the [C{\sc ii}] luminosity at fixed SFR is related to the strength of Ly$\alpha$ emission \citep[e.g.][]{Knudsen2016,Smit2017}, as [C{\sc ii}] emission may be reduced in ISM conditions that favour high Ly$\alpha$ escape. Indeed, conditions that may lower the strength of [C{\sc ii}] emission are typically found in Ly$\alpha$ emitters. Compared to the general galaxy population, strong Ly$\alpha$ emitters at $z\approx2-3$ are observed to have low metallicities and high ionisation states \citep[e.g.][]{Erb2016,Nakajima2016,Trainor2016,Kojima2017}. Our results however do not agree with such scenario, as CR7 has an extremely high Ly$\alpha$ luminosity and EW, but has [C{\sc ii}] luminosities similar to those expected for the local relation (both integrated, and for individual clumps). Therefore, the strength of Ly$\alpha$ emission may not be the most important property driving the scatter in the SFR-L$_{[\rm CII]}$ relation. Hence, indications of a potential relation between Ly$\alpha$ strength and [C{\sc ii}] luminosity may have reflected a more fundamental correlation. 
 
\begin{figure}
	\includegraphics[width=8.7cm]{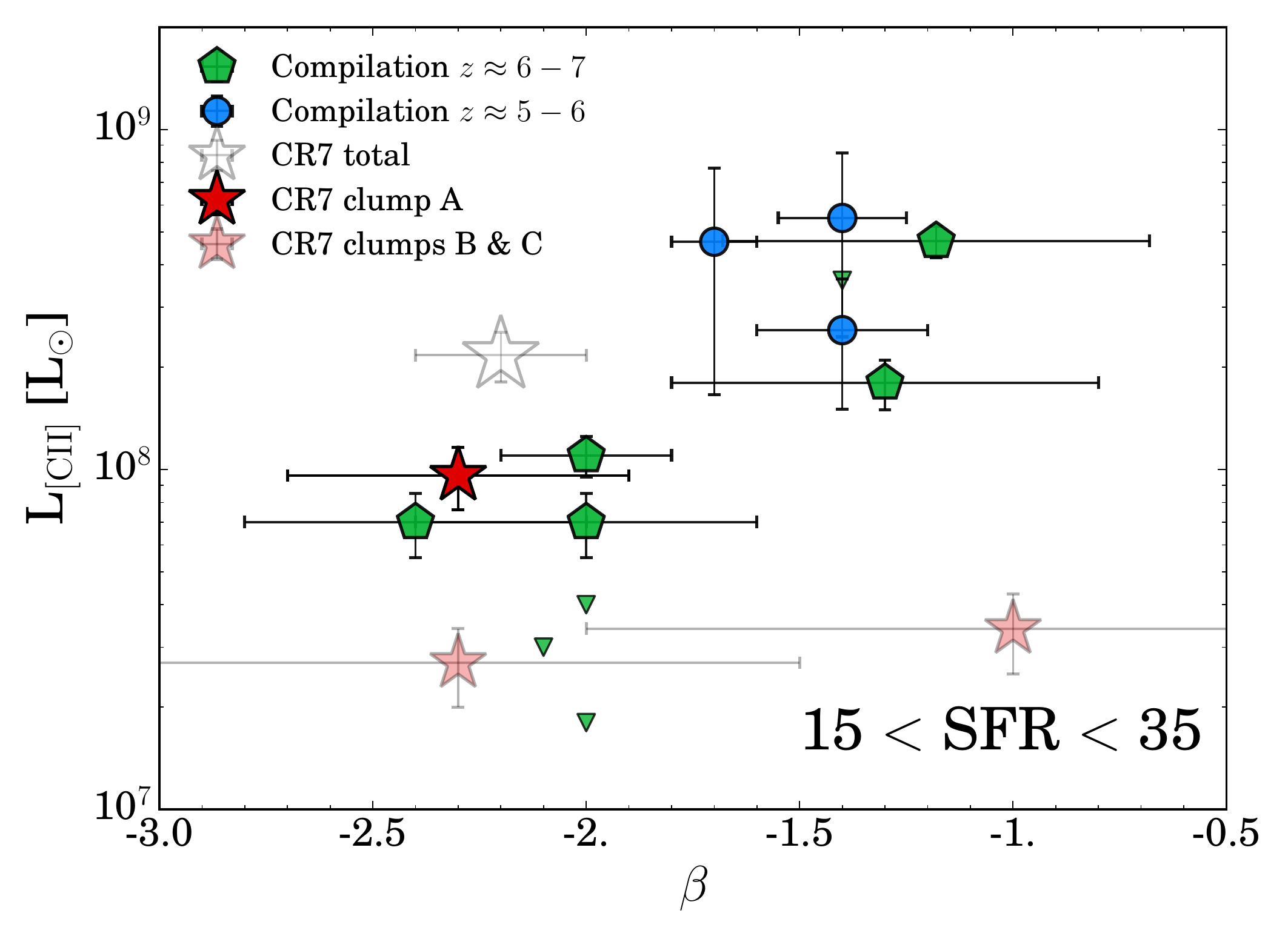}
    \caption{The relation between [C{\sc ii}] luminosity and UV slope, $\beta$, for galaxies with SFRs between $15-35$ M$_{\odot}$ yr$^{-1}$. Downward pointing triangles show galaxies with upper limits on the [C{\sc ii}] luminosity. Galaxies with redder colours are typically more evolved (older, higher metallicity), and consequently have higher [C{\sc ii}] luminosities. For reference, we show the measurements of clumps B, C and the total of CR7 in fainter colours.}
    \label{fig:beta_CII}
\end{figure}

A property that is related to the strength of Ly$\alpha$ emission in star-forming galaxies across $z\approx0-5$ is the UV slope, with bluer galaxies being brighter in Ly$\alpha$ \citep[e.g.][]{Atek2008,Matthee2016,Oyarzun2017}. The UV slope $\beta$ is closely related to the age, metallicity and dust attenuation law \citep[e.g.][]{Bouwens2012,Duncan2015,Mancini2016,Narayanan2017}, and therefore indirectly with properties that determine the [C{\sc ii}] luminosity at fixed SFR. We illustrate this in Fig. $\ref{fig:beta_CII}$, where we show the [C{\sc ii}] luminosity as a function of the UV slope for galaxies from the literature with SFR between 15 and 35 M$_{\odot}$ yr$^{-1}$. This is the range of SFRs in Fig. $\ref{fig:SFR_CII}$ that includes most galaxies at $z\approx5-7$ that have currently been observed. We note that we explicitly investigate a relatively narrow range in SFR to prevent being diluted by relations between SFR and galaxy properties themselves (for example, it has been established that fainter galaxies have bluer colours, \citealt{Bouwens2012}). Although the error-bars on UV slope measurements are large\footnote{Note that for the sample at $z\approx5-6$, we used the most recent UV slopes as measured by \cite{Barisic2017}, which differ significantly from the ground-based estimates from \cite{Capak2015}.}, it can be seen that at fixed SFR, galaxies with redder colours (which are typically more evolved) have higher [C{\sc ii}] luminosities. Since $\beta$ depends on several parameters, direct measurements of these parameters (such as metallicity) are required to fully understand the origin of this trend, and which parameter is most important.

It is possible to compare our results in more detail to the results obtained for Himiko in \cite{Ouchi2013}. Himiko consists of three UV components, with similar total UV and Ly$\alpha$ luminosity as CR7. Himiko has similar limits on its dust mass, but has not been detected in [C{\sc ii}], with a limiting luminosity L$_{\rm [CII]} \lesssim 5\times10^7$ L$_{\odot}$ \citep{Ouchi2013}, which is well below the total [C{\sc ii}] luminosity from CR7. However, unlike CR7, the individual components in Himiko have SFRs of $\approx 7, 5$ and 8 M$_{\odot}$ yr$^{-1}$ \citep{Ouchi2013}, and are hence likely of lower mass than CR7s clump A. The SFRs of Himiko's components are comparable to CR7 clumps B and C, that have lower [C{\sc ii}] luminosities than the existing [C{\sc ii}] limits for Himiko (and the clumps in Himiko are all very blue, with $\beta \lesssim -2.0$). Deeper [C{\sc ii}] observations at the resolution we obtained for CR7 could perhaps be able to detect those clumps. Another explanation could be that Himiko has a lower dynamical mass, such that supernova explosions disperse [C{\sc ii}] emitting gas over larger areas, decreasing its observability.

\begin{figure}
	\includegraphics[width=8.6cm]{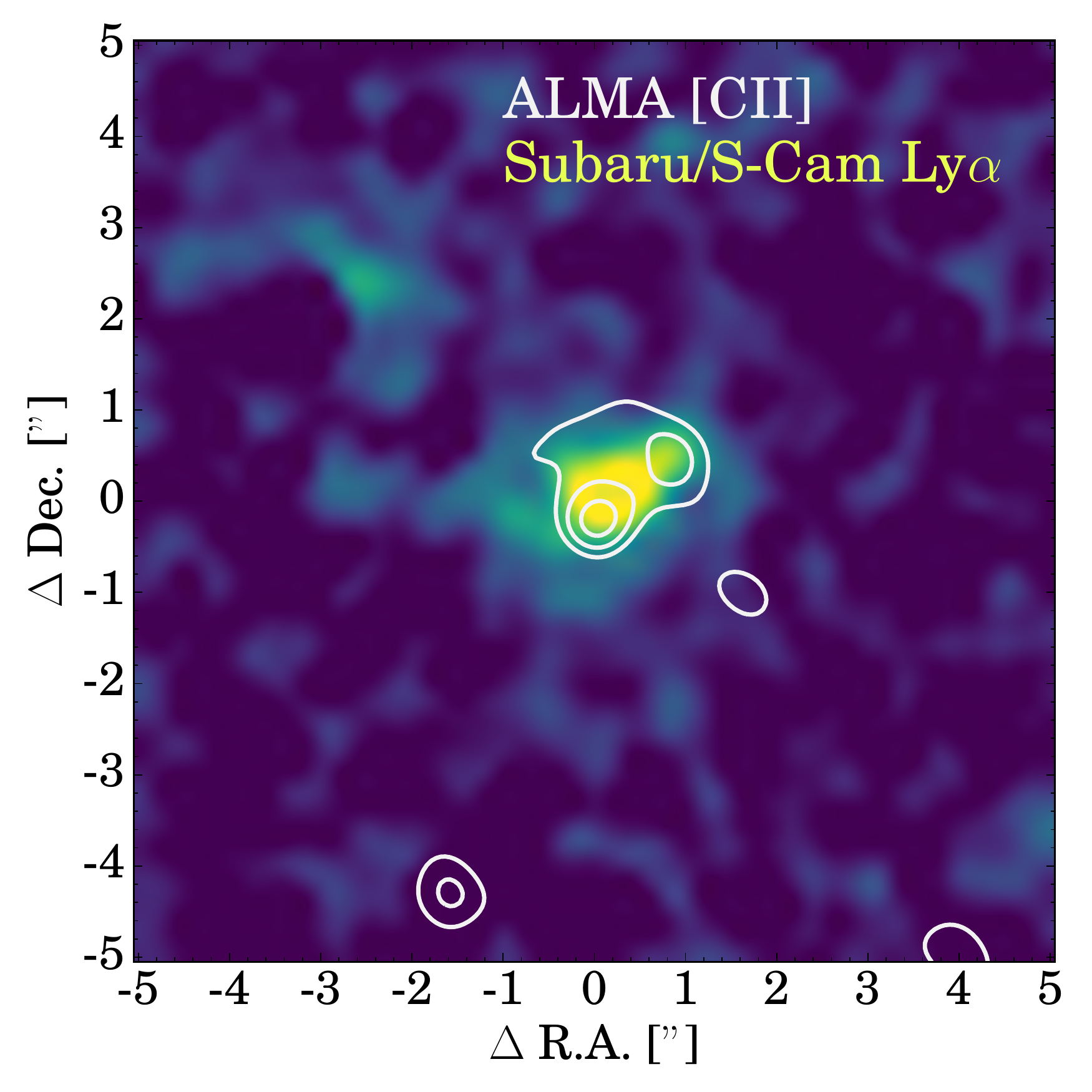}
\caption{Rest-frame Ly$\alpha$ image (obtained by subtracting the (weak) contribution from the continuum in the NB921 image using the Subaru/S-Cam $z'$ band image.) of CR7, overlaid with ALMA [C{\sc ii}] contours, with a scale of $54\times54$ kpc. The Ly$\alpha$ surface brightness is closely related to the [C{\sc ii}] surface brightness. At clump A, Ly$\alpha$ is redshifted with respect to [C{\sc ii}] by $127\pm27$ km s$^{-1}$.  \label{fig:ALMA_Lya}}
\end{figure}

\subsection{Spatial and spectral Ly$\alpha$-[CII] connection in CR7}
As shown in Fig. $\ref{fig:ALMA_Lya}$, the [C{\sc ii}] emission in CR7 coincides spatially with the Ly$\alpha$ emission. This may not be unexpected if both Ly$\alpha$ and [C{\sc ii}] originate from star-formation. Although [C{\sc ii}] is also emitted from gas that is (partly) neutral (due to its slightly lower ionisation energy than hydrogen; for example photo-dissociating regions), Ly$\alpha$ emission scatters in these neutral gas regions, such that it may also be observed from these regions. 

Interestingly, the [C{\sc ii}] line-width and velocity offset of clump A are surprisingly similar to the best fitted-shell model to CR7's Ly$\alpha$ profile \citep{Dijkstra2016}, which has an intrinsic line-width of 259 km s$^{-1}$ and outflow velocity of 230 km s$^{-1}$ (compare to Table $\ref{tab:measurements}$). Their best fitted shell-model assumes a negligible dust content (consistent with our constraints on the IR luminosity), and the resulting shell-model parameters are similar to Ly$\alpha$ emitters and analogue galaxies at lower redshifts. Therefore, the resemblance of the shell-model parameters based on the Ly$\alpha$ profile with the observed velocity offsets indicates feedback processes are already present in CR7. Further detailed studies on how the Ly$\alpha$ and [C{\sc ii}] line-profiles are related spatially require high resolution Ly$\alpha$ IFU observations with e.g. MUSE, as our current Ly$\alpha$ spectral information is limited to clump A.

\section{Discussion} \label{discussion}
\subsection{Multiple velocity components: merger/satellites?} \label{discuss_velo}
As shown in Figures $\ref{fig:velocity_2ds}$ and $\ref{fig:velocities}$, CR7 consists of several different components of [C{\sc ii}] emission, at different velocities. Similar clumpy structures are also observed in [O{\sc iii}]$_{88\mu \rm m}$ and [C{\sc ii}] emission in the $z=7.1$ galaxy BDF3299 \citep{Carniani2017}, and also in [C{\sc ii}] emission in a relatively massive galaxy at $z=6.1$ \citep{Jones2017}. Such relatively complex [C{\sc ii}] structures and the observed velocity offsets are also observed in hydro-dynamical simulations of galaxies at $z\approx6-7$ with halo masses of $\sim10^{11}$ M$_{\odot}$ \citep[e.g.][]{Vallini2013,Pallottini2017}. The line-profiles do not show evidence for broad-line components due to outflows \citep[e.g.][]{Gallerani2016}. Therefore, we are likely observing satellites (B, B-2, C, C-2) falling into a more massive central galaxy. These satellites have velocity widths similar to those of satellites in simulations by \cite{Pallottini2017b,Pallottini2017}, although simulated satellites typically have lower masses than those estimated for CR7's clumps. Clump A is the component that has the highest dynamical mass estimate (see Table $\ref{tab:measurements}$, although we stress that these are rough estimates) and likely the central galaxy. The dynamical mass estimate of clump C-2 is on the same order of magnitude, potentially indicating a major merger. Observations at higher resolution and with improved sensitivity are required to robustly resolve different components and perform more detailed kinematical/dynamical analyses.

\subsection{On the nature of CR7}
Here, we investigate what our new measurements mean for the interpretation of CR7. For a more in depth discussion on the updated detection significance of He{\sc ii} and limits on other high-ionisation UV lines we refer to \cite{Sobral2017}. The main conclusion from \cite{Sobral2017} is that the most luminous component of CR7 is likely undergoing a recent burst of star-formation, with no clear evidence for AGN activity and no convincing detection of UV metal lines in clump A. 

Initially, \cite{Sobral2015} argued that different UV components in CR7 were in different evolutionary stages, with the majority of the stellar mass possibly being found in clumps B and C (see also \citealt{Agarwal2016}). This interpretation was mostly based on the different UV slopes, with clump A being the bluest and hence the youngest. However, there are two uncertainties related to this interpretation. The first is that UV slopes are relatively uncertain (see Table $\ref{tab:measurements}$), and thus consistent with being the same for all sources. The second is that it is challenging to estimate where the majority of stellar mass is from {\it Spitzer}/IRAC observations, which has a relatively large PSF. For example, contrary to \cite{Sobral2015}, \cite{Bowler2016} found that the majority of stellar mass is likely associated to clump A using a deconvolution of the IRAC data based on UV detections as a prior. The new observations point towards a scenario where most of the mass is in clump A as the dynamical mass estimate of clump A is higher than that of clumps B and C. Simultaneously, the constraints from the UV slope, IR continuum and the Ly$\alpha$ luminosity indicates that clump A is also the youngest. 

As the SFR-[C{\sc ii}] ratio of different clumps are similar, there is evidence that all clumps have similar metallicities. However, [C{\sc ii}] luminosity may be reduced due to photo-evaporation by far-UV ($6-13.6$ eV) and ionising ($>13.6$eV) photons emitted from young stars in the vicinity, which typically happens a few tens of Myrs after the burst of star-formation \citep[e.g.][]{Decataldo2017,Vallini2017}, which can affect different clumps differently. The metallicities that we would infer from [C{\sc ii}] ($\approx0.1-0.2$Z$_{\odot}$) are inconsistent with the metallicity inferred by \cite{Bowler2016}, which is $\approx0.005$Z$_{\odot}$. However, their result depends on a He{\sc ii} strength that is now ruled out \citep{Sobral2017}. Based on upper limits on UV metal lines, the SFR and photoionization modelling, \cite{Sobral2017} inferred a metallicity of $\approx0.05-0.2$ Z$_{\odot}$, which is consistent with the metallicity indicated by the [C{\sc ii}] luminosity. Therefore, it is ruled out that the luminosity in all clumps in CR7 is dominated by PopIII-like stars, but note that small amounts of PopIII-like stars in unpolluted pockets of gas could still be present \citep[e.g.][]{Pallottini2015}. The fact that the SFR-[C{\sc ii}] ratio is similar to other $z\approx6-7$ galaxies and galaxies in the local Universe also indicates that photo-evaporation does not play a major role in lowering the [C{\sc ii}] luminosity. 

Due to its extreme Ly$\alpha$ luminosity $(\approx 10$L$^{\star}$, \citealt{Matthee2015}), CR7 could be powered by an AGN, particularly as the AGN fraction of Ly$\alpha$ emitters with similar luminosities at $z=2-5$ approaches 100 \% \citep{Calhau2017}. However, these galaxies typically have (much) broader Ly$\alpha$ lines and C{\sc iv} luminosities exceeding the observational limits for CR7 \citep[e.g.][]{Sobral2017b}, such that they are not fully comparable. The data presented in this paper do not favour an AGN explanation, particularly for clump A, because AGN typically have much lower log$_{10}$(L$_{\rm [CII]}$/L$_{\rm IR}$) ratios \citep[e.g.][]{Ota2014}. Furthermore, \cite{Bowler2016} show that the UV emission in clump A is slightly extended, which could also be at odds with an AGN-dominated scenario. Therefore, it is likely that clump A in CR7 is powered by a burst of star-formation with moderately low metallicity. It is remarkable that there is no UV nor IR continuum emission observed at the position of the ALMA detection C-2, which has a similar dynamical mass estimate as clump A. If potential UV emission is obscured, the dust is likely at a higher temperature (such that our IR continuum luminosity limit would be weak).

In order to improve our understanding of the CR7 system and other similar systems at $z\sim7$, it is most important to obtain more accurate constraints on the stellar masses, ages and metallicities of different clumps. While deeper {\it HST} imaging results in more strongly constrained UV slopes, the majority of progress will be made with integral field spectroscopic observations in the IR which can trace the rest-frame optical. Future adaptive optics assisted observations with MUSE can further map out the relation between the Ly$\alpha$ and [C{\sc ii}] kinematics, and simultaneously probe the large scale environment on $\sim 0.3-7$ Mpc scales.

\section{Conclusions} \label{sec:conclusions}
In this paper, we have presented deep follow-up observations of the CR7 galaxy in band 6 with ALMA targeted at the [C{\sc ii}] fine-structure cooling line and the FIR dust continuum emission (on source integration time 6.0h). After the discovery of CR7 as the most luminous Ly$\alpha$ emitter at $z\sim7$ \citep{Matthee2015}, the spectroscopic confirmation of Ly$\alpha$ \citep{Sobral2015}, the evidence for a narrow and strong He{\sc ii} line and the existence of several clumps seen in the {\it HST}/WFC3 imaging \citep{Sobral2015}, numerous spectacular interpretations emerged. While a high He{\sc ii}/Ly$\alpha$ ratio is now ruled out for clump A \citep{Sobral2017}, our ALMA observations provide independent measurements of the properties of the structure of the ISM around the three UV clumps observed in CR7 and furthermore reveal the presence of metals in its ISM. Our main results are:

\begin{enumerate}
\item We detect extended [C{\sc ii}]$_{\rm 158 \mu m}$ line-emission around CR7, with a total luminosity of $(2.17\pm0.36)\times10^8$ L$_{\odot}$ (Fig. $\ref{fig:ALMA}$). The [C{\sc ii}] emission is observed over an 2.2 arcsec$^2$ area with a typical blue-shift of $-167\pm27$ km s$^{-1}$ with respect to Ly$\alpha$, consistent with simple shell-model parameters based on an outflowing shell of low column density hydrogen. A fraction of  $0.72\pm0.18$ of the [C{\sc ii}] flux is associated with separate [C{\sc ii}] emitting clumps, see \S $\ref{sec:CII}$. 
 
\item Within CR7, four separate components can be identified in the three dimensional [C{\sc ii}] data-cube, see \S $\ref{CII_HST}$. UV components A and B are associated to clumpy [C{\sc ii}] emission at a similar blue-shift of $\approx 130$ km s$^{-1}$ with respect to the Ly$\alpha$ redshift, see Table $\ref{tab:measurements}$. A separate [C{\sc ii}] component, C-2, is blue-shifted by $\approx 100$ km s$^{-1}$ with respect to the systemic redshift. This component is not clearly associated to a UV clump, but closest to clump C. Around clump B we detect another [C{\sc ii}] component blue-shifted by $\approx 300$ km s$^{-1}$ compared to the systemic. 

\item The observed rich dynamical structure resembles the structure found in recent hydrodynamical simulations of $z\sim6-7$ galaxies with halo masses $\sim10^{11}$ M$_{\odot}$, and leads to the interpretation that these components are likely inflowing satellites. The measured line-widths and dynamical mass estimates (Table $\ref{tab:measurements}$) indicate that clump A contains most of the mass and is likely the central galaxy of the halo, although the dynamical mass estimate of clump C-2 is on the same order of magnitude, potentially indicating a major merger.

\item We do not detect FIR continuum emission, resulting in a 3$\sigma$ upper limit of L$_{\rm IR} < 2.8\times10^{10}$ L$_{\odot}$ (under the assumption that the dust temperature is 35 K), corresponding to M$_{\rm dust} < 2.7\times10^6$ M$_{\odot}$, see \S $\ref{sec:FIR}$. The limiting FIR-to-[C{\sc ii}] ratio indicates a low dust-to-metal ratio, much higher than typical sub-mm and quasar host galaxies. We use the FIR constraints to mitigate large uncertainties in the SFRs (due to uncertainties in their UV slopes) and find SFRs of $28^{+1}_{-1}$, $5^{+2}_{-1}$ and $7^{+1}_{-1}$ M$_{\odot}$ yr$^{-1}$ for clumps A, B and C, respectively.

\item Based on spatially resolved [C{\sc ii}] measurements, we compare the positions of clumps A, B and C and CR7 in total on the SFR-L$_{\rm [CII]}$ diagram with the literature, see \S $\ref{sec:SFRLCII}$ and Fig. $\ref{fig:SFR_CII}$. We find that all clumps and the total CR7 luminosity are consistent with the local relation (and other recent detections at $z\approx6-7$) within the scatter. Comparison with hydrodynamical simulations (and with the caveat in mind that poorly constrained ionisation conditions may play an important role), these measurements indicate gas metallicities of $0.1<$ Z/Z$_{\odot}<0.2$.

\item We find that for galaxies with SFRs between 15 and 35 M$_{\odot}$ yr$^{-1}$, the [C{\sc ii}] luminosity increases strongly with increasing UV slope, which is a tracer of dust, age and metallicity (Fig. $\ref{fig:beta_CII}$). This explains why several Ly$\alpha$ emitters are offset towards lower [C{\sc ii}] luminosities, as the Ly$\alpha$ escape fraction increases towards bluer UV slopes. 
\end{enumerate}

The picture that emerges is that in CR7 we are likely witnessing the build up of a central galaxy in the early Universe through complex accretion of satellites. The strong Ly$\alpha$ emission is likely powered by a star-burst with young, metal poor stars. Major improvements will be made with resolved spectroscopy that can measure gas metallicities and ionisation states, which can be used to constrain the properties of stellar populations in the different components.

\acknowledgments

We thank the referee for their constructive comments which have helped improving the quality and clarity of this work. We thank Raffaella Schneider for comments on an earlier version of this paper. We thank Leindert Boogaard, Steven Bos, Rychard Bouwens and Renske Smit for discussions.
JM acknowledges the support of a Huygens PhD fellowship from Leiden University. DS acknowledges financial support from the Netherlands Organisation for Scientific research (NWO) through a Veni fellowship and from Lancaster University through an Early Career Internal Grant A100679. AF acknowledges support from the ERC Advanced Grant INTERSTELLAR H2020/740120. BD acknowledges financial support from NASA through the Astrophysics Data Analysis Program (ADAP), grant number NNX12AE20G. Based on observations made with ESO Telescopes at the La Silla Paranal Observatory under programme ID 294.A-5018. This paper makes use of the following ALMA data: ADS/JAO.ALMA\#2015.1.00122.S. ALMA is a partnership of ESO (representing its member states), NSF (USA) and NINS (Japan), together with NRC (Canada) and NSC and ASIAA (Taiwan) and KASI (Republic of Korea), in cooperation with the Republic of Chile. The Joint ALMA Observatory is operated by ESO, AUI/NRAO and NAOJ. 

%

\vspace{5mm}
\facilities{ALMA (band 6), {\it HST} (WFC3), VLT (X-SHOOTER), Keck (DEIMOS), VISTA (Vircam), Subaru (S-Cam).}






\bibliography{CR7_ALMA.bib}



\end{document}